\documentclass[nohyper,11pt,letterpaper]{JHEP3}
\usepackage{epsfig}
\usepackage{graphicx}

\DeclareSymbolFont{AMSa}{U}{msa}{m}{n}
\DeclareSymbolFont{AMSb}{U}{msb}{m}{n}
\let\Box\relax
\DeclareMathSymbol{\Box}{\mathord}{AMSa}{"03}

\def \eqn#1#2{\begin{equation}#2\label{#1}\end{equation}}

\title{Vacuum Bubble in an Inhomogeneous Cosmology}

\author{W. Fischler, S. Paban, M. \v{Z}ani\'{c}\\
Department of Physics \\ University of Texas \\ 1 University Station, C1608 \\ Austin,
TX 78712 \\
E-mail: \email{fischler@physics.utexas.edu, paban@physics.utexas.edu, marija@mail.utexas.edu}}

\author{C. Krishnan \\ International Solvay Institutes \\ Physique 
Th\'{e}orique et Math\'{e}matique \\ ULB C.P. 231, Univerisit\'{e} Libre 
de Bruxelles, \\ B-1050, Bruxelles, Belgium \\
E-mail: \email{Chethan.Krishnan@ulb.ac.be}}

\abstract{We study the propagation of bubbles of new vacuum in a radially
inhomogeneous Lema\^{i}tre-Tolman-Bondi background that
includes a cosmological constant.  This exemplifies   
the classical evolution of a tunneling bubble through a metastable state 
with 
curvature inhomogeneities, and  
will be relevant in the context of the 
Landscape. We demand that the matter profile in the LTB background 
satisfy the weak energy condition. For sample profiles 
that satisfy this restriction, 
we find that the evolution of the 
bubble (in terms of the physically relevant
coordinates intrinsic to the shell) is largely unaffected by 
the prsence of local inhomogeneities. Our setup 
should also be a useful toy model for capturing 
the effects of ambient inhomogeneities on an inflating region. 
}


\received{???????? ?st, 2000} \accepted{???????? ?th, 1998}
\preprint{\\UTTG-10-07\\ULB-TH/07-31}

\begin{document}




\section{\bf Introduction}

Although bubble propagation in homogeneous cosmological backgrounds
has been amply studied \cite{Israel:1966rt}-\cite{Lake:1984pn}, bubble 
propagation in
inhomogeneous backgrounds is virtually unexplored 
\cite{Khakshournia:2002jr}. 
Since tunneling is of enormous importance \cite{Vilenkin:1983xq, 
Hawking:2006ur} in populating the string Landscape \cite{Kachru:2003aw, 
Susskind:2003kw, Aguirre:2007an, Chang}, we expect that the evolution of a 
new Universe through 
an old one, is worthy of study.

One reason for the lack of emphasis on inhomogeneous 
backgrounds is the cosmic no-hair theorem \cite{Gibbons:1977mu, 
Hawking:1981fz, Starobinsky:1982mr}.
This theorem states that the evolution of an expanding universe that 
contains a cosmological constant will eventually be dominated by 
vacuum energy, and will asymptote to a de Sitter space. It will 
imply in particular, that inhomogeneous backgrounds containing a 
cosmological constant (like the ones constructed on Landscape vacua) will 
evolve toward 
`ironing out' such inhomogeneities, and therefore it will make sense to 
restrict the study of bubble propagation to homogeneous
backgrounds. 
But fluctuations can still clump locally even in an asymptotically de 
Sitter space, and therefore the question of 
evolution of vacuum bubbles in an inhomogeneous spacetime is still 
relevant. In the context of eternal inflation and the string landscape, we 
have to face up to the possibility that the Universe that we inhabit is a 
result of bubble evolution through an inhomogeneous relic Universe. In 
principle this could give rise to experimental signatures in the form of 
CMB fluctuations, about which we comment in the final section. 

One particular work that stands out in this context is that of Wald 
\cite{Wald:1983ky}, who did a comprehensive study of of 
the late time evolution of anisotropic but homogeneous spaces with a 
cosmological constant. In this restricted setting, he was able to 
define the conditions for the cosmic no hair theorem to be satisfied by 
taking advantage of the Bianchi classification of homogeneous, 
anisotropic metrics.  Wald studied the time evolution of such metrics 
when the energy momentum tensor is the sum of two components: a 
cosmological constant and a term that satisfies the dominant  and strong 
energy conditions. He found that with the exception of the 
Bianchi IX, the other models will always asymptote to a de Sitter space, 
within a time scale $\sqrt{\frac{3}{\Lambda}}$. In
the case of the Bianchi IX models, the future behavior depends on the 
relative sizes of the cosmological constant and the spatial-curvature. 
Only when the latter exceeds the cosmological constant will the time 
evolution {\em not} be asymptotically de Sitter.

  It is often claimed that inflation solves  the horizon and flatness 
problems but,  as already remarked in its
early years, these claims are based on the assumption that the  
pre-inflationary state of the universe can be described by a homogeneous 
and isotropic Friedmann-Robertson-Walker (FRW) space. Under these strong assumptions, inflation does indeed reduce considerably the amount of fine tuning  necessary to explain the current experimental measurements. In general, however,  one expects a high degree of inhomogeneity in the initial conditions, with some regions of the universe expanding while others contract.  Holographic cosmology  might be the only exception to this rule because it naturally produces a homogeneous and almost flat space-time   \cite{Banks:2001px, Banks:2003ta, Banks:2004cw}.

The 
question of whether an inflating region can
continue to inflate if the ambient region is inhomogeneous, has been 
explored by several authors for different 
pre-inflationary backgrounds.
In a set of papers  
\cite{Goldwirth:1989pr,Goldwirth:1991rj,Goldwirth:1989vz}, Goldwirth and Piran sought to answer the question of whether large inhomogeneity in the very early universe could prevent it from entering a 
period of inflation. In their work,
inflation was not driven by a cosmological constant but by an inflaton field. Their analysis is numerical and restricted to an inhomogeneous but isotropic and  closed-universe background. They found that a large initial inhomogeneity could indeed suppress the onset of inflation, and that  the inflaton field must have a suitable value over a region of several horizon sizes in order for inflation to start.  This last finding raises the issue of an acausal initial condition. 
Other numerical simulations disagree with these results claiming that inhomogeneous backgrounds will have enough inflation to explain current observations \cite{Kurki-Suonio:1993fg,Berera:2000xz}. 
Analytical computations that use the long wave-length approximation seem to confirm that large inhomogeneities of the spatial curvature prevent the onset of inflation \cite{Deruelle:1994pa, Iguchi:1996jy, Iguchi:1996rh}. 

In this work, we will not be able to conclusively settle this issue once
and for all. Instead, our goal here is to study a particular problem that
we believe captures some of the same physics by playing local curvature
effects against the might of the cosmological constant. The problem we
consider is the classical evolution of a vacuum bubble in an
inhomogeneous background, containing a cosmological constant and dust.
This background, which we 
will refer to as the Lema\^{i}tre-Tolman-Bondi (LTB)  space-time  
\footnote{ We slightly generalize it from the original form to allow for 
a positive cosmological constant.}, will be the simplest possible 
inhomogeneous space, spherically symmetric and with only one center 
\cite{Lemaitre:1933gd}-\cite{Bondi:1947av}. The bubble will be created in 
a region of space that is expanding but will encounter through its 
evolution regions of varying curvature. In a nutshell, the problem is 
that the critical size for bubble nucleation depends locally on the scale 
factor. As this factor changes as a function of the position of the 
shell, it is possible for a supercritical bubble to become subcritical  
\footnote {
A bubble that can become subcritical in the future is a fluctuation, not 
a phase transition. }. Our analysis will be confined to the evolution of 
the bubble, it will not address the interesting issue of tunneling 
probabilities in inhomogeneous backgrounds. This latter issue is very 
complicated, some attempts to address it have
 been made in \cite{Abbott:1987xq}. 

Before embarking on describing the details, we emphasize that we are 
{\em not} 
dealing with the most general inhomogeneous bubble evolution imaginable. 
There are three major assumptions we make to render the problem tractable, 
and we list them below. The results of this paper should therefore be 
regarded as exploratory, rather than conclusive.
\begin{itemize}
\item The first assumption we make is that the ambient spacetime into 
which the bubble evolves is the radially inhomogeneous LTB form. This 
allows us to work with an exact solution of Einstein's equation, but the 
restricted form of the metric and matter in this solution implies that we 
are not dealing with the most general radially inhomogeneous cosmology 
admissible. But as we explain in more detail in the next section, this 
still leaves us with a great deal of flexibility in choosing the 
inhomogeneity profiles. On the inside of course, we are dealing with a 
vacuum bubble, 
so we take the matter there to be only a cosmological constant without 
loss of generality.
\item The second assumption that we make is that the evolution of the new 
phase can be adequately captured using a thin-wall approximation. In 
particular this means that we can deal with all the transient phenomena 
between the two regions using a thin layer. The more general scenario 
would be to model the matter leakage from one region 
to the other using a gradual profile, but we believe the 
thin-wall captures the phenomena involved to a first 
approximation.  
\item The above two assumptions together give us some traction. Once we 
make them, the 
problem that we need to solve reduces to that of 
solving the Israel junction conditions for a thin shell. But the junction 
conditions do 
not fully fix the evolution of the shell until we specify an equation of 
state on the bubble-wall. In the purely general relativistic context that 
we are working in, these equations of state are not fixed by the 
dynamics. So what we do 
here is to take the first step in this direction by considering a few 
equations of state of the form $p= w \rho$, for a range of values of $w$. 
We make some comments elsewhere on how one might try to fix this equation of 
state using field-theoretic arguments, but the full solution of this issue 
in the complicated background that we are working in, is likely to be 
difficult. In any event, as a start, we settle here for an 
exploration of these equations of state, leaving a more thorough analysis 
for future.

\end{itemize}

We investigate a few different scenarios using our numerical formalism. 
The first is the homogeneous limit of LTB, namely 
Robertson-Walker cosmologies with or without matter. We reproduce and 
extend known results in the literature using our methods, and offer 
some 
comments and new perspectives. Particular attention is paid to the case of 
bubble-propagation in an FRW universe with a Big Crunch in the future. We 
find that there are new ways (details are in a later section) 
in which the bubble can evolve in this case: it can even recollapse. The 
final scenario we consider is an LTB space that is truly 
inhomogeneous. In our examples, we pick our inhomogeneity profiles to 
have a fluctuation that is as sharp as possible while still respecting 
the weak energy condition. Surprisingly, 
we find that the qualitative bubble 
evolution (as seen in the physically relevant coordinates intrinsic to the 
shell or in the inside coordinates) is essentially the same with or 
without the fluctuation. This suggests that the energy conditions work 
towards the smoothening out of exotic bubble evolution patterns.

In the next section we shall describe the LTB model that represents the background on which the bubbles will propagate. In section 3, we will review the propagation of bubbles in a homogeneous background. We will consider the already known case where the outside background is simply a de Sitter space, either flat or closed, and the novel case where matter is added to the cosmological constant. In this latter case, we shall explore the interesting situation of a bubble moving in a background that evolves from expanding to collapsing. 
In section 4 we shall study the truly inhomogeneous 
LTB background. 
Our results, future directions and cautionary remarks are put together in
the Conclusion.
We have relegated to the Appendix, an explanation of 
the known formalism to study the propagation of bubbles in general relativity.

\section{\bf Setup: The Bubble and the Background} 

A propagating bubble divides the space-time into three regions: outside, shell and inside. 
To give a proper description of this space-time in general relativity we 
will make use of the junction conditions, first presented by Israel  
\cite{Israel:1966rt} and extensively used since then by many authors. In particular, we will follow the implementation of this formalism developed by Berezin, Kuzmin and Tkachev \cite{Berezin:1987bc}.

The model we will study consists of a bubble of true vacuum propagating on 
a metastable state whose energy-momentum tensor contains a higher 
cosmological 
constant  and dust. 
The bubble is assumed to be a thin-shell, with a perfect-fluid energy-momentum tensor whose equation of state will be allowed to vary. The true vacuum will be assumed homogeneous and isotropic while the outside background will be assumed to be spherically symmetric about one point in space. 
The line element of the \underline{outside region} \cite{Peebles:1967} is: 

\eqn{metric_b}{ds^2=d{t}^2 - \frac{[a(t,r) + r 
a'(t,r)]^2}{1-\frac{r^2}{R^2(r)}} dr^2 - a^2(t,r) r^2 d\Omega_2}
The expansion parameter $a(t,r)$ is both a function of time and of the radial coordinate $r$. 
Partial derivatives with respect to the radial coordinate will be represented by a prime. The function
$R(r)$ is an arbitrary function of the radial coordinate only and will be taken to be positive everywhere.
We will study the evolution of the bubble for different choices of  $R(r)$. It can be seen from these equations that we can choose a function 
$R(r)$ and
then integrate for $a(t,r)$ at each $r$. Then we can use this $a(t,r)$ to
{\em
define} the $d(t,r)$ in (\ref{EOMB}), with the caveat that the
choice of $R(r)$ has to be made in such a way that this energy density
needs to be positive.  
Therefore, the description of the spacetime is 
essentially complete (apart from fixing certain initial conditions which 
we will get to in a minute), once we specify the function $R(r)$. Because 
it is our prejudice that curvature will interfere with the expansion
of the bubble when its value is comparable to the cosmological constant,
we will choose $R(r)$ such that it can locally overwhelm the effect of 
the cosmological constant.

The equations of motion for the expansion factor and the dust density $d(t,r)$, in units where $8 \pi G=c=1$, are:

\begin{eqnarray} 
\Big(\frac{\partial_t a(t,r)}{a(t,r)}\Big)^2 + \frac {1}{a^2(t,r) R^2(r)} 
& = &  \frac{A} {a^3(t,r)} + \frac{\Lambda_{out}}{3}  \label{EOM}\\
\frac{1}{3} d(t,r) a^2(t,r) ( a(t,r) + r a'(t,r)) & = & A  \\ \nonumber \label{EOMB}
\end{eqnarray}
$A$ is a constant. These equations reduce to the familiar Friedmann-Robertson-Walker equations in
the limit where $R(r)$ is independent of the radial coordinate. Besides choosing $R(r)$ the solution to the equations of motion will involve the choice of  the initial condition $a(t_0,r)$. For simplicity, we will 
choose this function to be independent of the radial coordinate $r$.

The results that will be presented in the forthcoming sections 
correspond
to different choices
of the function $R(r)$. 
 For different
choices of $R(r)$, the collapse-profile can be completely different. One
way to understand this is to notice that the Friedmann-type
equation here can be thought of as an energy equation for a particle
of zero total energy in a potential
\eqn{potential}{V(a)=-\frac{A}{a}-\frac{\Lambda_{out}
a^2}{3}+\frac{1}{R^2}}
For those values of $r$ for which the maximum of this potential is above
zero, if the universe
starts at a big-bang, it will recollapse, but for other values of $r$
there is no recollapse.

The \underline{bubble} will be assumed to be a thin-shell (described 
by a hypersurface $\Sigma$) and spherically symmetric. Its energy momentum 
tensor will be assumed to be of a perfect-fluid type: 
$ {S_\tau}^\tau= \sigma, {S_\theta}^\theta={S_\phi}^\phi= -P$, $P=w 
\sigma$  and with the metric:

\eqn{metricbubble1}{{ ds^2|}_{\Sigma}=  d \tau^2  - \rho^2(\tau) 
d\Omega_2 
}
The \underline{interior of the bubble} will be assumed homogeneous and described by 

\eqn{metricbubble2}{ ds^2 = dT^2 - b^2(T) \left( \frac{dz^2}{1+z^2} + z^2 
d\Omega_2 \right) }
with the following equation of motion for the scale factor:

\eqn{FRWbubble}{ \left( \frac{db}{dT}  \right)^2 = 
\left( \frac{\Lambda_{in}}{3} \right) b^2(T) + 1}
We are assuming the inner space-time of the bubble to be open as  derived from a tunneling process \cite{Coleman:1980aw}.

Given this setup, our aim is to understand the evolution of the bubble. 
The dynamics of the bubble is governed by the Israel junction conditions, 
the details of which are relegated to an appendix. The coupled 
differential equations that govern shell evolution turn out to be 
(\ref{sigmaequation}), (\ref{Ton}), or in a more explicit form, 
(\ref{eomforshell1}), (\ref{eomforshell2}).

As always in general relativity, along with the dynamical equations, we 
also need an equation of state to fully specify the dynamics. 
Determining the equation of state between $\sigma$ and $P$ on the 
shell from first principles is a difficult problem that requires a field 
theoretic model for matter on either side. Since the LTB metric for a 
generic choice of the curvature profile $R(r)$ is known only numerically, 
this is doomed from the start. In the absence of a detailed understanding 
of this phase transition we will confine our study to perfect-fluid 
shells, $P = w \, \sigma$, and explore various values of $w$. 

It turns out that in the absence 
of dust on the outside, the shell-evolution equations can be written 
entirely 
in terms of 
$\rho(\tau)$ because (\ref{Ton}) becomes trivial, but 
this is no longer possible when dust is included. Indeed, equation 
(\ref{sigmaequation}) can be written as (reproduced from the appendix):
\begin{eqnarray}
\dot{\rho}^2 & = & -1 + B^2 \rho^2,  \label{EOMR}\\ \nonumber \\
B^2 & = & \frac{\Lambda_{in}}{3} + \left(  \frac{\sigma}{4} +  
\frac{1}{\sigma} \left( \frac{\Lambda_{out}-\Lambda_{in}}{3} + 
\frac{A}{a^3} \right) \right)^2. 
\end{eqnarray}
In general, to fix the evolution of $B(\tau)$ we need 
(\ref{Ton}), along with a knowledge of the scale factor through numerical 
solution of (\ref{EOM}). 

We also get constraints on the parameters and the allowed phase space 
from the positivity of energy, in particular, the condition that $\sigma$ 
be positive. The explicit relations that we will use later can again be 
found in the appendix.

\section{\bf Bubble Expansion in a Homogeneous Background}

The study of the bubble evolution can be done in different sets of coordinates. When dust is present, the problem of interest is most conveniently analyzed in the coordinates of the outside background.  To familiarize ourselves with these coordinates  we will devote this section to analyze
a simpler problem. The simplification will come from restricting the outside background to be homogenous
($R(r) = \mbox{constant}$). First, we will redo the well studied motion of a vacuum shell in a cosmological constant
background. Then we will add a dust energy density to the cosmological constant.

\subsection{Homogeneous Background without Matter}

The Israel junction conditions (\ref{sigmaequation})-(\ref{Ton}) tell 
us that 
it is consistent to have a vacuum shell (surface energy is constant and $w=-1$) in a transition that 
separates two de Sitter spaces with different values of the cosmological constant. Since this problem
has been solved many times in the literature we will only quote the results \cite{Berezin:1987bc} . 

In terms of the intrinsic bubble coordinates, the motion of the shell is given by:

$$ \rho(\tau) = \frac{1}{B} \cosh {B \tau} $$
where 

$$ B^2 =  \frac{\Lambda_{in}}{3} + \left( \frac{\sigma}{4} + \frac{\Lambda_{out}-\Lambda_{in}}{3 \sigma} \right) ^2 $$
regardless of the curvature of the outside background. The effect of this curvature only becomes manifest
when we express the motion of the bubble in terms of the outside coordinates. 
In these coordinates the evolution of the position of the bubble is given by:

\eqn{xevol}{ \frac{d x}{dt} = \frac{- (1- \frac{x^2}{R^2}) \sqrt{(\frac{\Lambda}{3} - \frac{1}{a^2 R^2})} \pm \sqrt{(B^2 a^2 x^2 -1)(B^2-\frac{\Lambda}{3})(1- \frac{x^2}{ R^2})}}{ a^2 x (B^2-\frac{1}{a^2 R^2})} }
where the  further assumption that $\Lambda_{in}=0$ and $\Lambda_{out}=\Lambda$ has been made. The evolution of the scale factor is given by the well known expression

$$ \Big(\frac{da/dt}{a}\Big)^2 + \frac{1}{a^2 R^2} = 
\frac{\Lambda}{3}  \label{FRWk}$$
For  times $ t >> \frac {1}{H} = \sqrt{3/\Lambda}$,  (\ref{xevol}) reveals the following asymptotic behavior :

\begin{eqnarray}
a(t) & \rightarrow &   \frac{1}{2 H  R} \,\, e^{H t} \hspace{4ex}  \mbox{when} \hspace{2ex}  a(0)=  \frac{1}{ H R}  \\ \nonumber \\
x(t)  &  \rightarrow &   R \sin \left( \frac{\mbox{constant}}{R} \right) - O( e^{- Ht}) \hspace{4ex}  \mbox{if}  \hspace{1ex}  R>0 \hspace{1ex} \mbox{ and finite}  \label{x1} \\ \nonumber\\ 
x(t) & \rightarrow &    \mbox{constant}+ O(e^{-Ht})  \hspace{6ex}  \mbox{if}  \hspace{1ex}  R \rightarrow \infty  \label{x2} \\  \nonumber \\
\frac{dx}{dt} & \rightarrow & \frac{1}{a} \,\, \sqrt{1 - \frac{x^2}{R^2}} 
\, \, 
\frac{ \frac{\sigma}{4}  - \frac{\Lambda}{ 3 \sigma}}{\frac{\sigma}{4}  + \frac{\Lambda}{ 3 \sigma}} \label{sol}
\end{eqnarray}

\begin{figure}
\centering
\hspace{-0.8cm}\begin{minipage}[b]{.5\textwidth}
\centering
\includegraphics[width=6cm]{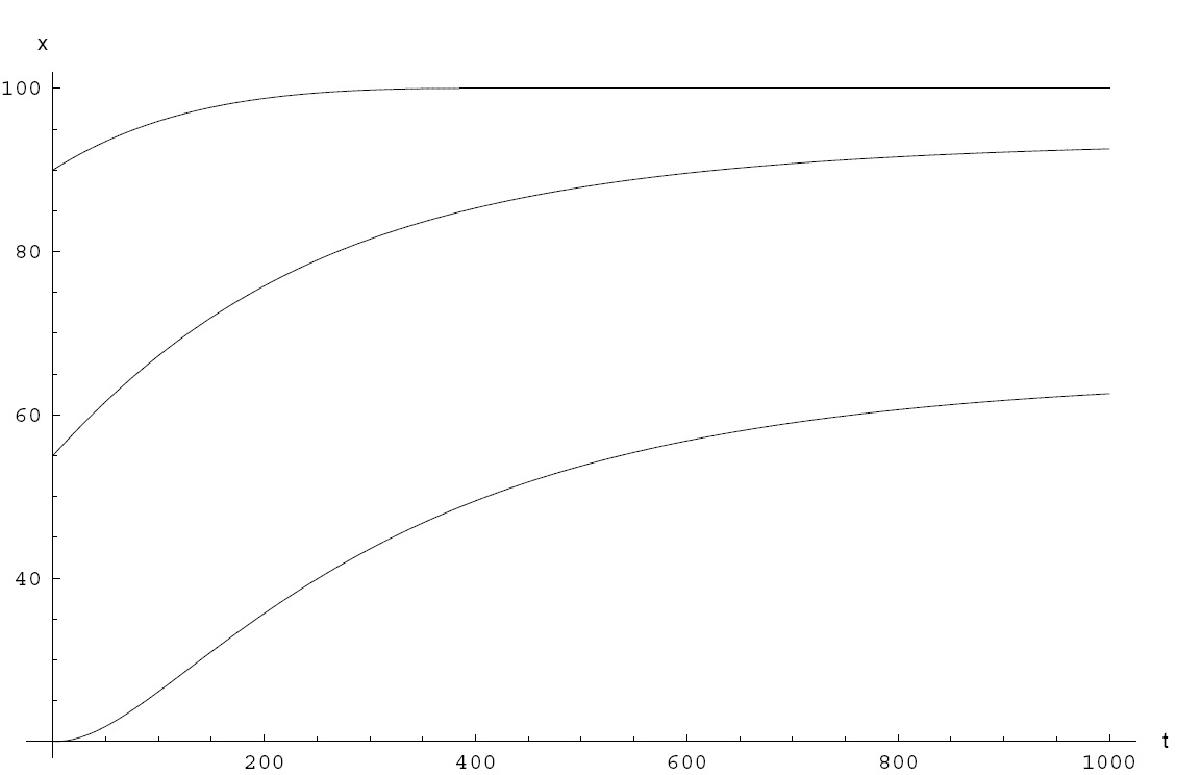}
\end{minipage}%
\begin{minipage}[b]{.55\textwidth}
\centering
\includegraphics[width=6cm]{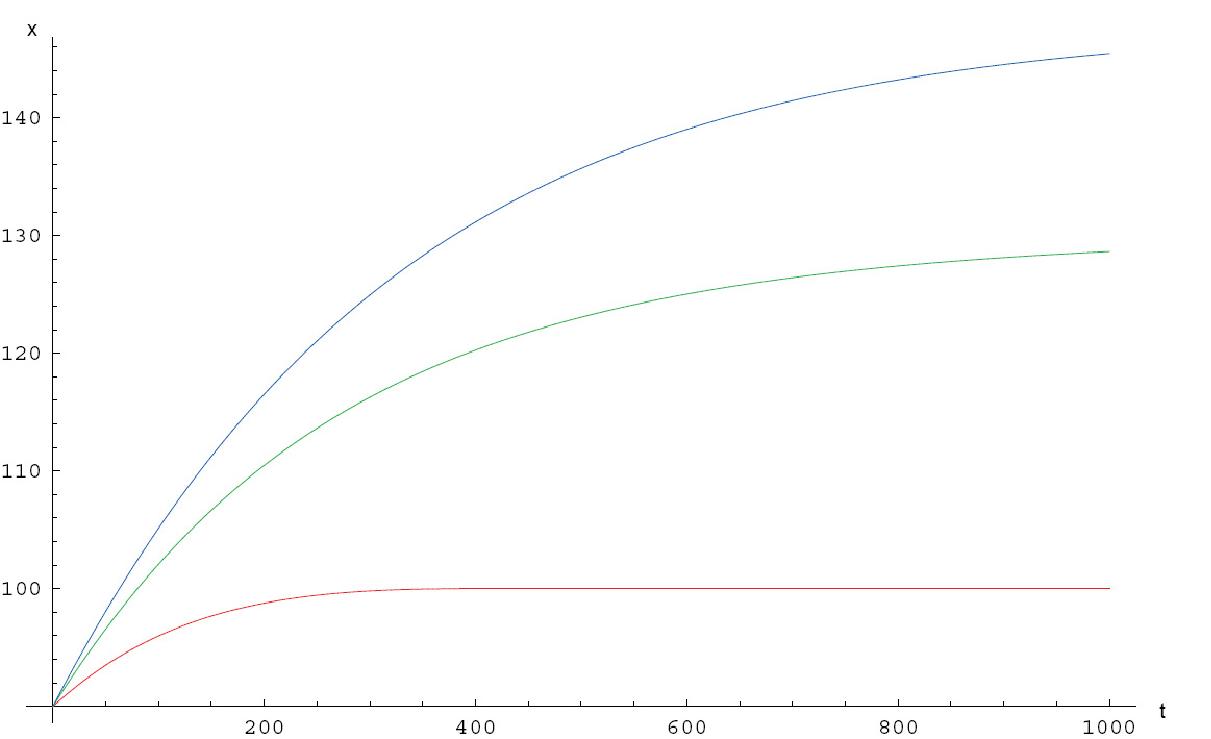}
\end{minipage}\\[-10pt]
\hspace{-1.4cm}\begin{minipage}[t]{.4\textwidth}
\caption{
Time evolution of the bubble in  the outside coordinates, $x[t]$, for several initial sizes,
$x_{init}=20,55,90$. The background is assumed homogeneous ($R=100$) and without matter.
The scales on the axes depend on the parameter values we have chosen and 
are therefore not terribly important. 
}\label{figure_x_vac_nom}
\end{minipage}%
\hspace{1.8cm} \begin{minipage}[t]{.4\textwidth}
\caption{
Time evolution of the bubble in the outside coordinates, $x[t]$, in homogeneous backgrounds without matter.
Colors correspond to different choices of R; $R=100$ red, $R=150$ green, $R=\infty $ blue.
} \label{figure_x_diffR_vac_nom}
\end{minipage}%
\end{figure}

The value of the constants in (\ref{x1}) and ( \ref{x2}) is very dependent on initial conditions as can be seen in Figure \ref{figure_x_vac_nom}. Also (\ref{sol})  corroborates the well known fact \cite{Coleman:1980aw} that bubbles propagate asymptotically in time at the speed of light in the thin wall limit of $\Lambda \ll 3 \sigma $.

 A space with positive curvature only makes sense if $a(0) \geq \frac{1}{RH} $. Once the initial value exceeds this critical value the space expands forever. The effect of the curvature becomes
eventually negligible. Thus it is not surprising that the evolution of the bubble is similar regardless of the curvature. 
There is a difference in the possible asymptotic value that $x$ can take; in the positive curvature case
$x$ always has to remain below $R$. The rate of
the expansion is asymptotically given by $B$ in the shell coordinates and by $H$ in the
outside coordinates. 
In terms of the inside coordinates the motion of the bubble is given by:

$$ Z(T) = \frac{1}{B} \sqrt{1 + B^2 ( T - T_0)^2}$$
This expression can be derived from the metric matching condition at the 
shell (\ref{mink-match1}) and 
assuming that $b(T_0)=1$, which is a good assumption when $\Lambda_{in}=0$. $T_0$ is the inside time at which the bubble is created. For late times the bubble propagates at the speed of light as expected \cite{Coleman:1980aw}. Ultimately, we are always interested in the motion of the bubble as viewed by the observer inside the bubble. The moral of this example is that the motion, as viewed by the inside observer,  is the same regardless of the amount of curvature outside the bubble. 

\begin{figure}[htp]
\centering
\includegraphics [width = 12 cm]{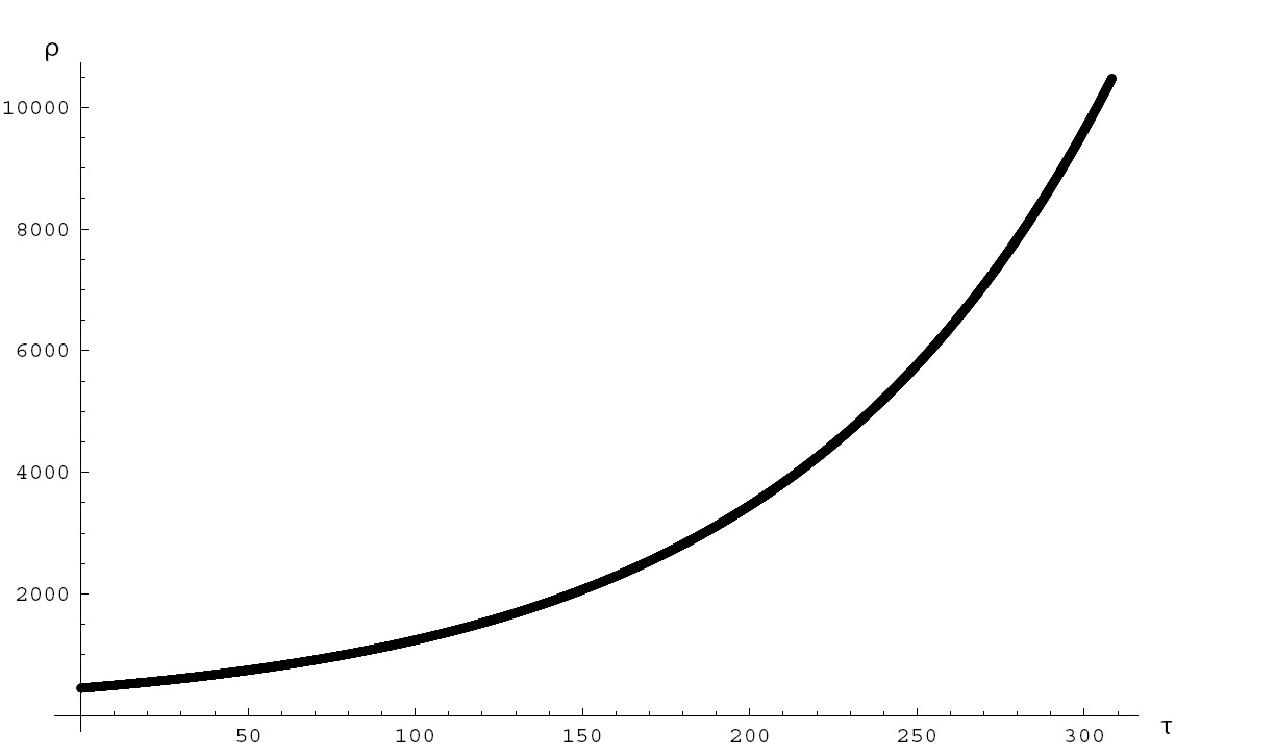}
\caption{Time evolution of the bubble in the coordinates of the bubble, $\rho[\tau]$, in homogeneous backgrounds without matter 
(same as in Figure \ref{figure_x_diffR_vac_nom}).}
\label{figure_rho_diffR_vac_nom}
\end{figure} 

In our equations above (and in the ones that follow) we have chosen the
natural units $c=8\pi G=1$. Since we are dealing only with classical
evolutions, $\hbar$ never comes up, so we have the freedom to pick one
more unit. In the numerical evolution that we will undertake later, this
means that there is rescaling of the coordinates (by a dimensionful
parameter) that we are still free to do. 
We will choose the various parameters (like $\Lambda$, $\sigma$, $A$) 
within a few orders of magnitude: the actual values themselves are not too 
relevant because of the
above-mentioned freedom in choosing the scale for the coordinates.
Physically we expect  the cosmological constant to
reflect the scale corresponding to the minimum of some effective scalar
potential, whereas the bubble stress-tensor should reflect  the
potential barrier between the two minima. 
For a generic potential, we expect that the
scales involved will be not too different in order of magnitudes. As far 
as the evolution of the shell goes, the only place where we
could have trans-Planckian effects becoming significant is when there is a
singularity in the spacetime itself. In particular the fact that the
shell surface is an actual physical wall, and {\em not} a surface at
conformal
infinity (like a horizon), means that we do not have to worry about finite
energy modes observed far away being red-shifted trans-Planckian modes
\cite{Farhi:1986ty, farhi2, fischler1, fischler2}.

The effects of curvature on the bubble propagation, as seen by the outside observer,  are illustrated in Figures \ref{figure_x_vac_nom},
 \ref{figure_x_diffR_vac_nom} and \ref{figure_rho_diffR_vac_nom}. To simulate these evolutions we have chosen the following values for the parameters, in units of $8 \pi G=1$:
\begin{eqnarray}
 \Lambda & =  &  3 \times 10^{-5}   \nonumber \\
 a_{init} & = & 5  \nonumber \\
\gamma_{out} & = & \gamma_{in}=+1 \nonumber \\
\sigma_{init} & = &10^{-3} 
\label{initial-conditions}
\end{eqnarray}
These values  for the parameters can not be chosen independently since they have to satisfy the condition  (\ref{xi_constraint}) that  yields:

\eqn{sigma_constr} {\sigma \leq 2 \sqrt { \frac {\Lambda} {3} } = 6.3 \times 10^{-3}}
Also from the equation (\ref{xevol}) the lower and the upper bound on $x$ are:

\eqn{x_range} { \frac {1}{aB} \leq x \leq R }
We have chosen the initial size of the bubble in accordance with this range.

In particular, Figure \ref{figure_x_vac_nom} shows the evolution of the bubble in
the outside coordinates for various initial sizes. Note that the maximum size of the bubble
in these coordinates corresponds to the limit $x=R$. 

Figure \ref{figure_x_diffR_vac_nom} shows the effects of curvature on bubble propagation
by comparing evolutions of bubbles with the same initial size $x_{init}$, but in different
curvature background. We see that more curvature corresponds to a slower evolution of the bubble 
in the outside coordinates; however, the corresponding bubble evolution $ \rho[\tau]$ 
in the bubble coordinates is exactly the same, irrespective of the amount of background curvature,
and it is shown in Figure \ref{figure_rho_diffR_vac_nom}.

\begin{figure}[htp]
\centering
\includegraphics [width = 12 cm]{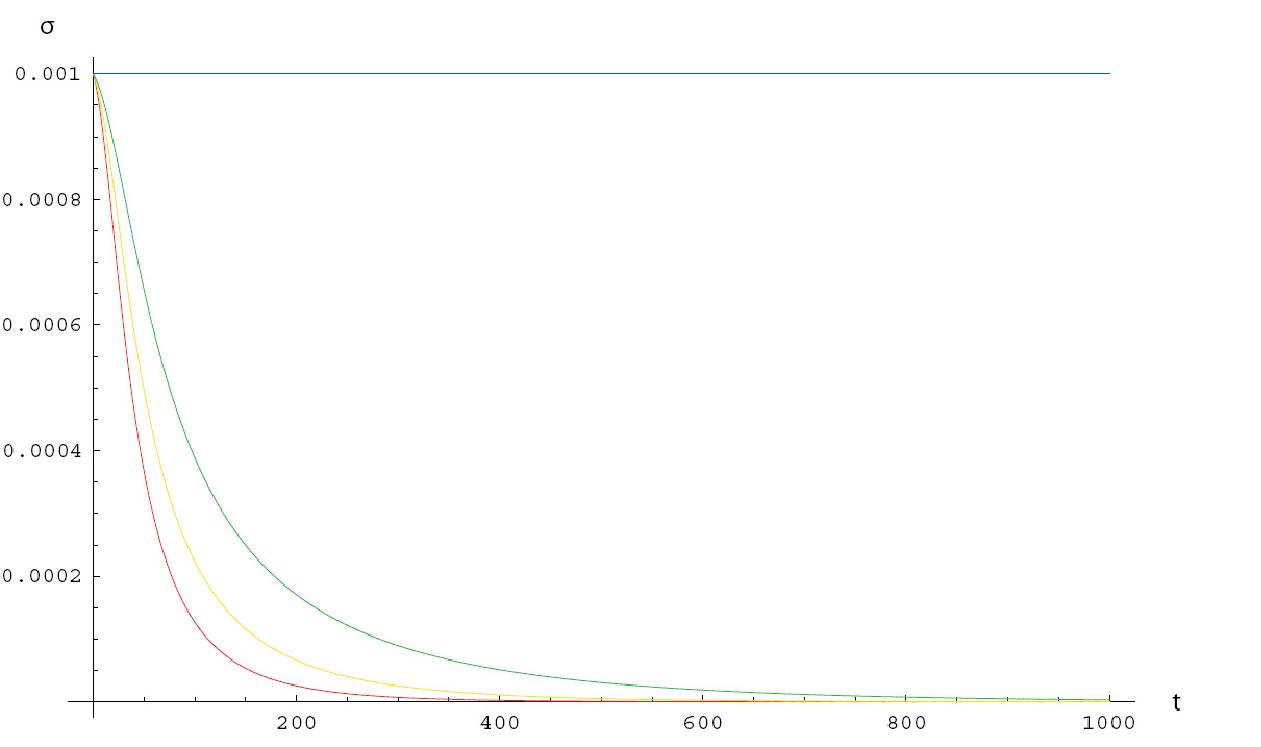}
\caption{Time evolution of the surface energy density $\sigma [t]$ on the bubble. The background is assumed 
homogeneous ($R=100$) and without matter.
Colors correspond to different equations of state; $w=1/3$ red, $w=0$ yellow, $w=-1/3$ green, $w=-1$ blue. }
\label{figure_sigma_fixedR100_nom}
\end{figure} 

\begin{figure}
\centering
\hspace{-0.8cm}\begin{minipage}[b]{.5\textwidth}
\centering
\includegraphics[width=6cm]{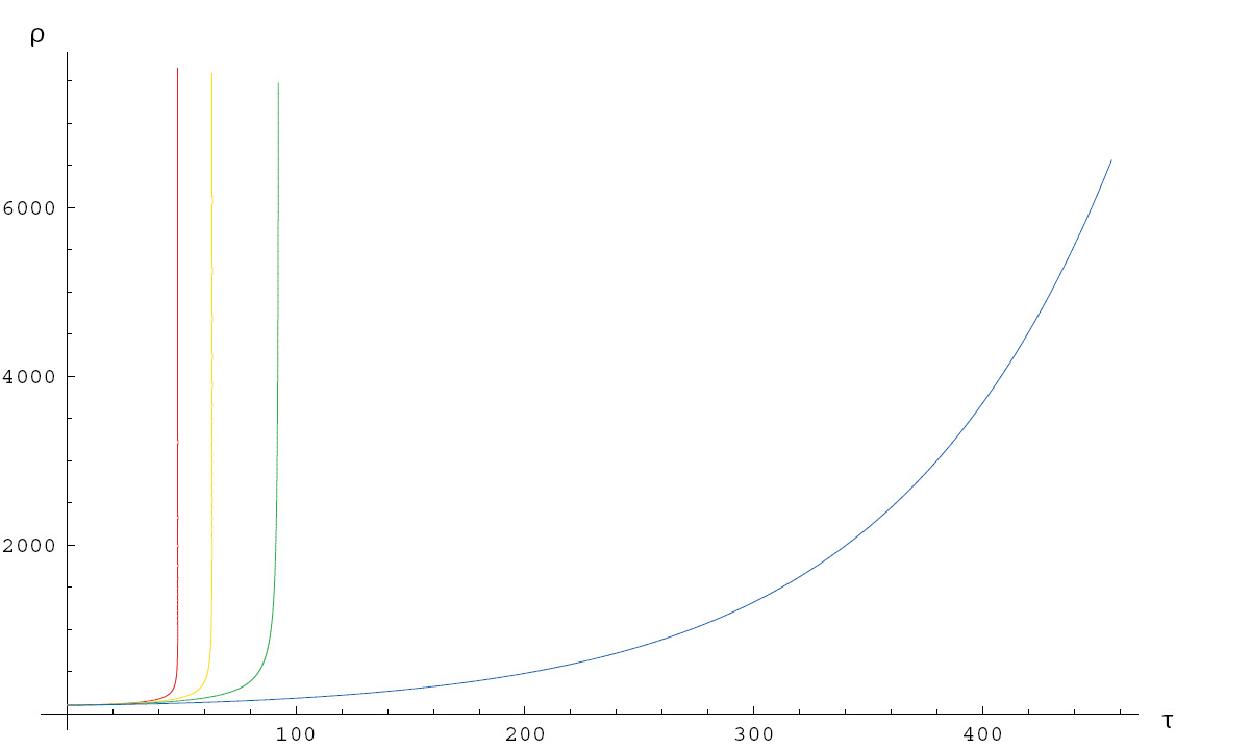}
\end{minipage}%
\begin{minipage}[b]{.55\textwidth}
\centering
\includegraphics[width=6cm]{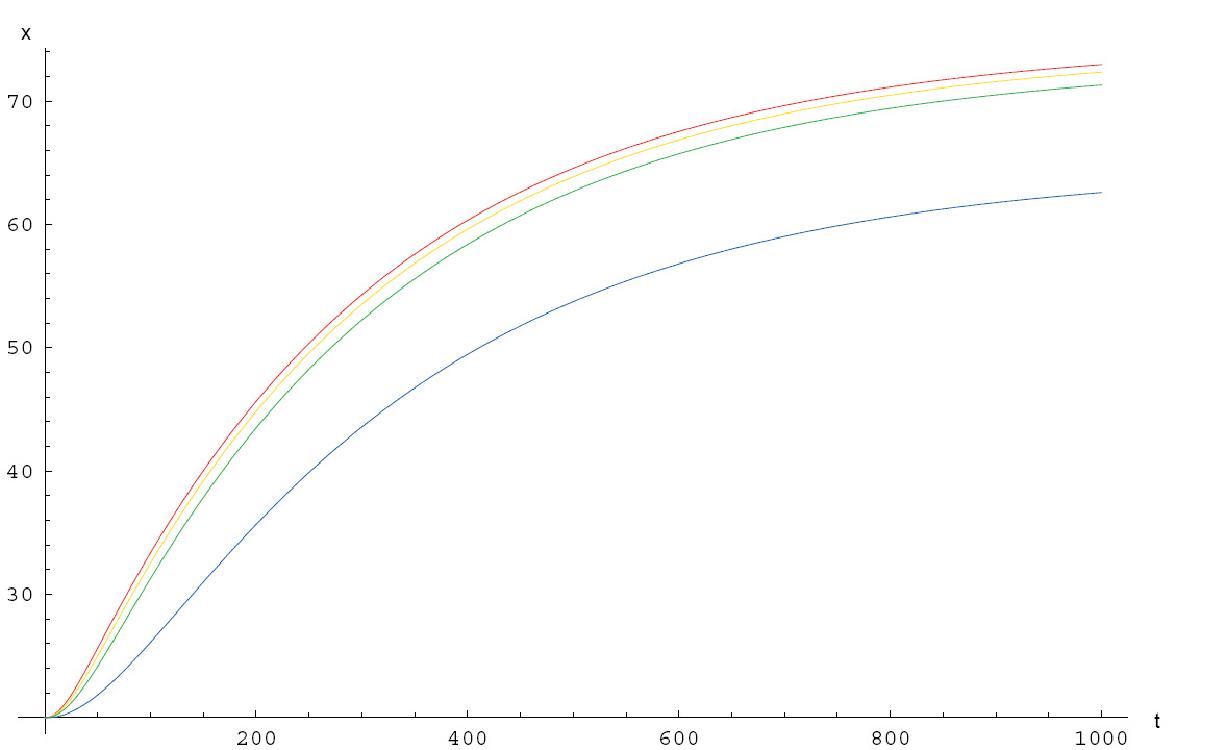}
\end{minipage}\\[-10pt]
\hspace{-1.2cm}\begin{minipage}[t]{.4\textwidth}
\caption{
Time evolution of the bubble in the coordinates of the bubble, $\rho[\tau]$.
The background is assumed homogeneous ($R=100$) and without matter.
Colors correspond to different equations of state; $w=1/3$ red, $w=0$ yellow, $w=-1/3$ green, $w=-1$ blue.
}\label{figure_rho_fixedR100_nom}
\end{minipage}%
\hspace{2cm} \begin{minipage}[t]{.4\textwidth}
\caption{
Time evolution of the bubble in the outside coordinates, $x[t]$. The background is assumed
homogeneous ($R=100$) and without matter. $x_{init}=20$.
Colors correspond to different equations of state; $w=1/3$ red, $w=0$ yellow, $w=-1/3$ green, $w=-1$ blue.
} \label{figure_x_fixedR100_nom}
\end{minipage}%
\end{figure}

In addition, one might wonder what would happen if we did not assume the 
bubble to be
a vacuum bubble, namely if $w\neq -1$. Then the evolution of the energy 
density $\sigma$ on the
bubble is governed by (\ref{eomforshell2}). In the absence of dust on the 
outside the solution to this equation is:

\eqn{sigmawnot}{\sigma = \frac{ \zeta}{\rho^{2(1+w)}} } 
where $\xi$ is a constant. When substituting this solution on 
(\ref{EOMR}) we obtain the following equation for $\rho(\tau)$

\eqn{rhoevolwnot}{ \dot{\rho}^2 = -1 +  \left( \frac{ \zeta}{ 4 \rho^{(1+2 w)}} + \frac{\Lambda}{3 \zeta}  \, \rho^{3+2w} \right) ^2 } 
For $\rho$  large, the solution to this equation takes the form:

\eqn{rhownot}{ \rho(\tau) \sim \frac{\rho(\tau_0)}{ \left[ 1 - (1+ 2 w) \frac{\Lambda}{3 \zeta} \rho(\tau_0) ^{1 +  2 w} ( \tau-\tau_0) \right] ^{1/(1+2 w)} } }
This analysis of the asymptotic behavior is also captured in the results plotted   in Figure \ref{figure_sigma_fixedR100_nom}, and Figure \ref{figure_rho_fixedR100_nom}.

Figure \ref{figure_x_fixedR100_nom} depicts the corresponding bubble evolution in the outside coordinates, $x[t]$. 

{\em In summary}, in this section we have considered a closed FRW 
background that will eventually asymptote to a de Sitter space.
The outside curvature doesn't affect the evolution of the vacuum bubble 
when it is described in terms of the shell coordinates, and therefore,
nor in terms of the inside coordinates. The 
curvature only makes a difference when the evolution is studied in terms 
of the outside coordinates. In these coordinates, the effect of a bigger 
curvature is to slow down the evolution of the comoving bubble 
coordinate. The asymptotic value of this coordinate depends on the 
initial condition but is always smaller that the curvature radius.  When 
the equation of state of the shell is not that of vacuum, the evolution 
in the shell coordinates is much faster than that of vacuum but is also 
independent of the outside curvature. When described from the point of 
view of the outside coordinates, this evolution is faster than the 
corresponding one for the vacuum shell, but is qualitatively similar.

\subsection{Homogeneous Background with Matter}

To study the evolution of the bubble in a space that might contract we have to modify the outside
energy density. One option is to introduce matter. In this instance the evolution of the scale parameter
is given by

$$ \Big(\frac{da}{dt}\Big)^2 +\frac{1}{R^2} - \frac{\Lambda}{3} a^2- 
\frac{A}{a}=0 
$$


If  $ A \leq \frac{1}{R^3} \sqrt{\frac{4}{9 \Lambda}}$ and $a(0)$ is small enough the expansion of the universe will reverse into contraction. With this new form of outside energy density, however, it is no longer consistent with the matching conditions to have a constant surface energy density on the bubble, not even when $\sigma+P=0$. The evolution will be given by the equations:

\begin{eqnarray}
\frac{dx}{dt} & = &{\frac{ - \left( 1 - \frac{x^2}{ R^2} \right) 
\frac{da/dt}{a} \pm \sqrt{ \left( 1- \frac{x^2}{ R^2} \right) (a^2 
B^2 x^2 
-1) \left(B^2 - \left(\frac{\Lambda}{3} + \frac{A}{a^3}\right) \right)} }{x a^2 ( B^2- \frac{1}{a^2 R^2})} \label{hom_x} } \\ \nonumber \\ 
\frac{d \sigma}{dt} & = & - 2 \left( \frac{da/dt}{a} + 
\frac{dx/dt}{x} \right)( \sigma + P ) + 
\gamma_{out}\frac{dx/dt}{\sqrt{- 
(dx/dt)^2 a^2 + 1 - \frac{x^2}{ R^2}}} \frac{3 A}{a^2} \label{hom_sigma} 
\\ 
\nonumber \\
B & = &  \frac{\sigma}{4} + \frac{1}{ \sigma} \left(\frac{\Lambda}{3}+ \frac{A}{a^3} \right) \label{hom_B}
\end{eqnarray}
As in the previous section in order to solve these equations an assumption about the equation of state on the bubble was needed. We did make the assumption that the shell is made of a perfect fluid with equation of state $ P = w \, \sigma$, and explored several values for $w$.
For the dust energy to be comparable to the cosmological constant at early times, we choose
\eqn{matter}{ A  = 10^{-4}}
in Planck units. The critical curvature which is the minimal curvature needed for the space to 
turn around and eventually collapse is given by:

\eqn{R_cr} {R_{cr} = \left( \frac {9 \Lambda A^2}{4} \right) ^{-1/6} = 107}
The position of the maximum of the potential $V[a]$ is given by:

\eqn{a_max} {a_{max}= \left( \frac{3A}{2\Lambda} \right)^{1/3} = 1.7}
In our simulations we choose the initial value of the scale factor $a$ to be $a_{init}=a(0) = 1$.

The outside background geometry will be completely specified once we choose
the amount of curvature. We will consider several cases.

\subsubsection {$R > R_{cr}$}

The universe will always expand. If at the instant the bubble is created $a(0)$ is to the left of max of the potential (\ref{potential}), then the universe will experience a period of slower growth until it eventually goes over the max and the expansion becomes dominated by the cosmological constant. 

We will now investigate bubble propagation on such background. We are again assuming
that the inside region has zero energy density, and that
$\gamma_{out}=\gamma_{in}=+1$. In this case, the condition (\ref{xi_constraint}) yields:

\eqn{sigma_constr} {\sigma \leq 2 \sqrt { \frac {\Lambda} {3} + \frac {A} {a^3} }}
This constraint will be the strongest as $a\rightarrow \infty $, when, as is the case without matter, it reduces to:

\eqn{sigma_lower} { \sigma \leq 6.3 \times 10^{-3}}
Once again, we choose:

\eqn{sigma_init} {\sigma_{init} = 10^{-3}}
where $\sigma_{init}$ is given in Planck units.
From the equation (\ref{hom_x}) the lower and the upper bound on $x$ are:

\eqn{x_range} { \frac {1}{aB} \leq x \leq R }

We will choose the initial size of the bubble in accordance with this range
and compare the evolution for several equations of state on the surface of the bubble.
Decreasing $R$ (and therefore increasing the curvature of the space) will have the effect
of slowing down the evolution, as long as $R > R_{cr}$.

\begin{figure}
\centering
\hspace{-0.8cm}\begin{minipage}[b]{.5\textwidth}
\centering
\includegraphics[width=6cm]{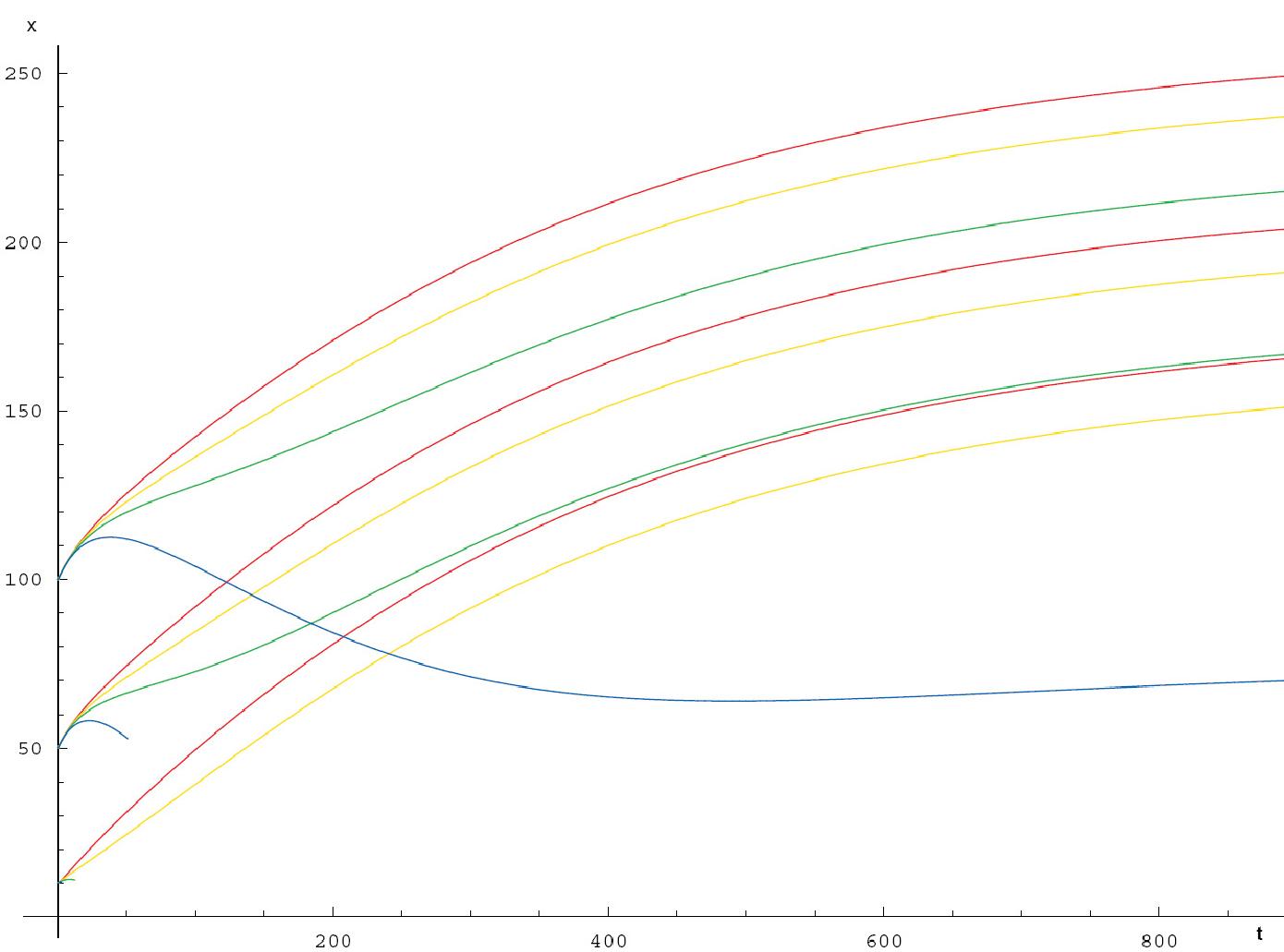}
\end{minipage}%
\begin{minipage}[b]{.55\textwidth}
\centering
\includegraphics[width=6cm]{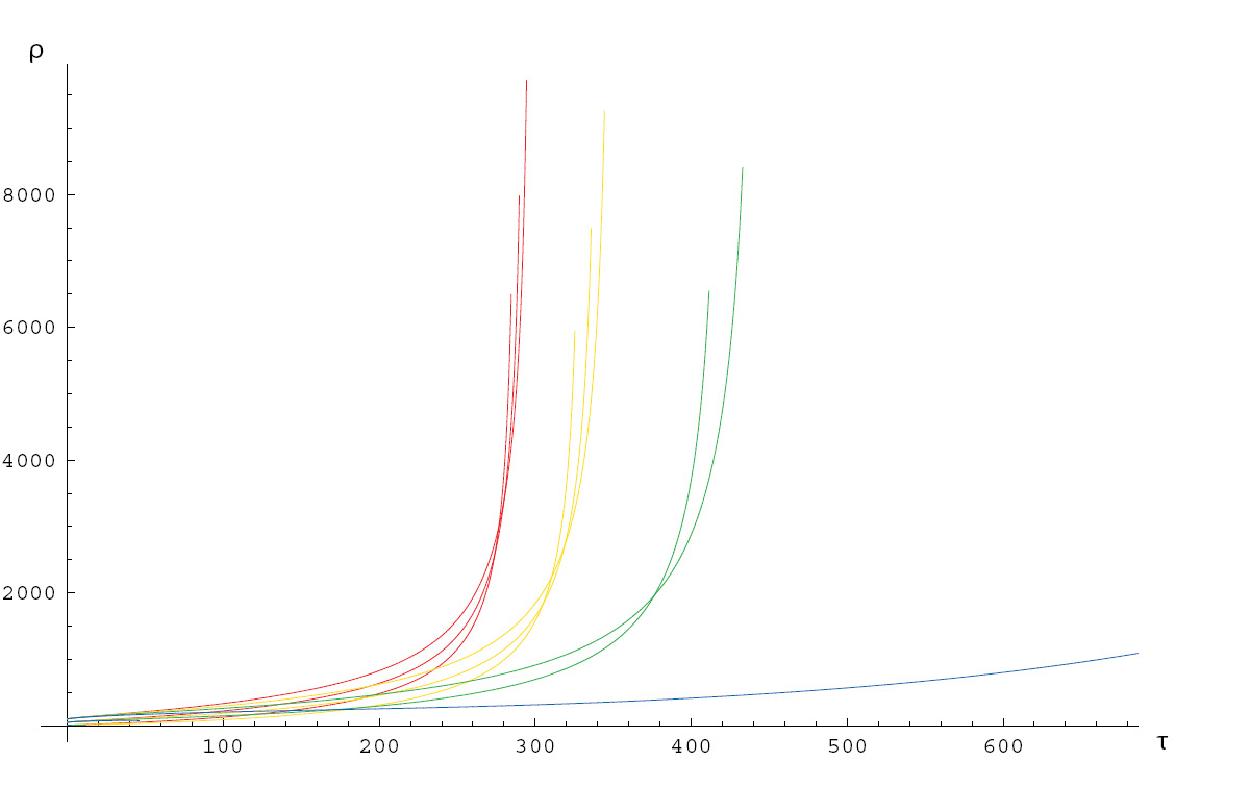}
\end{minipage}\\[-10pt]
\hspace{-1.2cm}\begin{minipage}[t]{.4\textwidth}
\caption{Time evolution of  the bubble in the outside coordinates,
$x[t]$, for several initial sizes, $x_{init}=10,50,100$. The background
contains homogeneous curvature ($R=500$) and matter
($A=10^{-4}$). Colors correspond to different equations of state;
$w=1/3$
red, $w=0$ yellow, $w=-1/3$ green, $w=-1$ blue.}\label{figure_x_fixed_500}
\end{minipage}%
\hspace{1.2cm} \begin{minipage}[t]{.4\textwidth}
\caption{Time evolution of the bubble in the bubble coordinates,
$\rho[t]$, for the same background conditions as in
Figure \ref{figure_x_fixed_500}.
Colors correspond to different equations of state; $w=1/3$ red, $w=0$
yellow, $w=-1/3$ green, $w=-1$ blue.
} \label{figure_rho_fixed_500}
\end{minipage}%
\end{figure}

First let us consider the case with very little curvature, $R=500$. Figure 
\ref{figure_x_fixed_500} shows the time evolution of bubbles of several
initial sizes in the outside coordinates. In most cases bubbles grow without
reaching the upper bound, $x=R$. Smaller $w$ slows down the expansion rate, and in particular,
the choice of $w=-1$ soon leads to a contracting bubble in the outside coordinates.
Depending on the initial size of the bubble, $w=-1$ bubbles either reach
the lower bound, $x=\frac{1}{aB}$, which results in the breakdown of the simulation,
or eventually stabilize and stop contracting.

Time evolution in the coordinates on the bubble (Figure \ref{figure_rho_fixed_500})
reveals behavior similar to the case without matter. Namely, for the bubbles that do survive,
the evolution is slower than in the case without matter, but the overall qualitative behavior 
stays the same.

\begin{figure}
\centering
\hspace{-0.8cm}\begin{minipage}[b]{.5\textwidth}
\centering
\includegraphics[width=6cm]{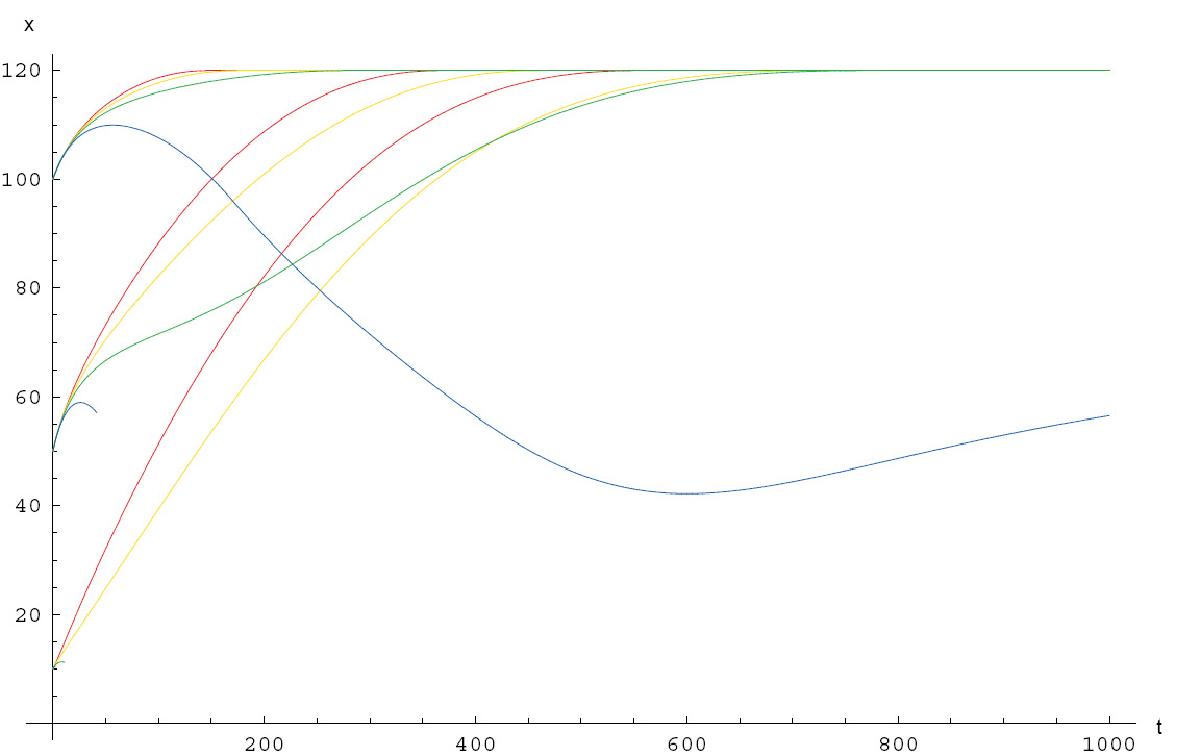}
\end{minipage}%
\begin{minipage}[b]{.55\textwidth}
\centering
\includegraphics[width=6cm]{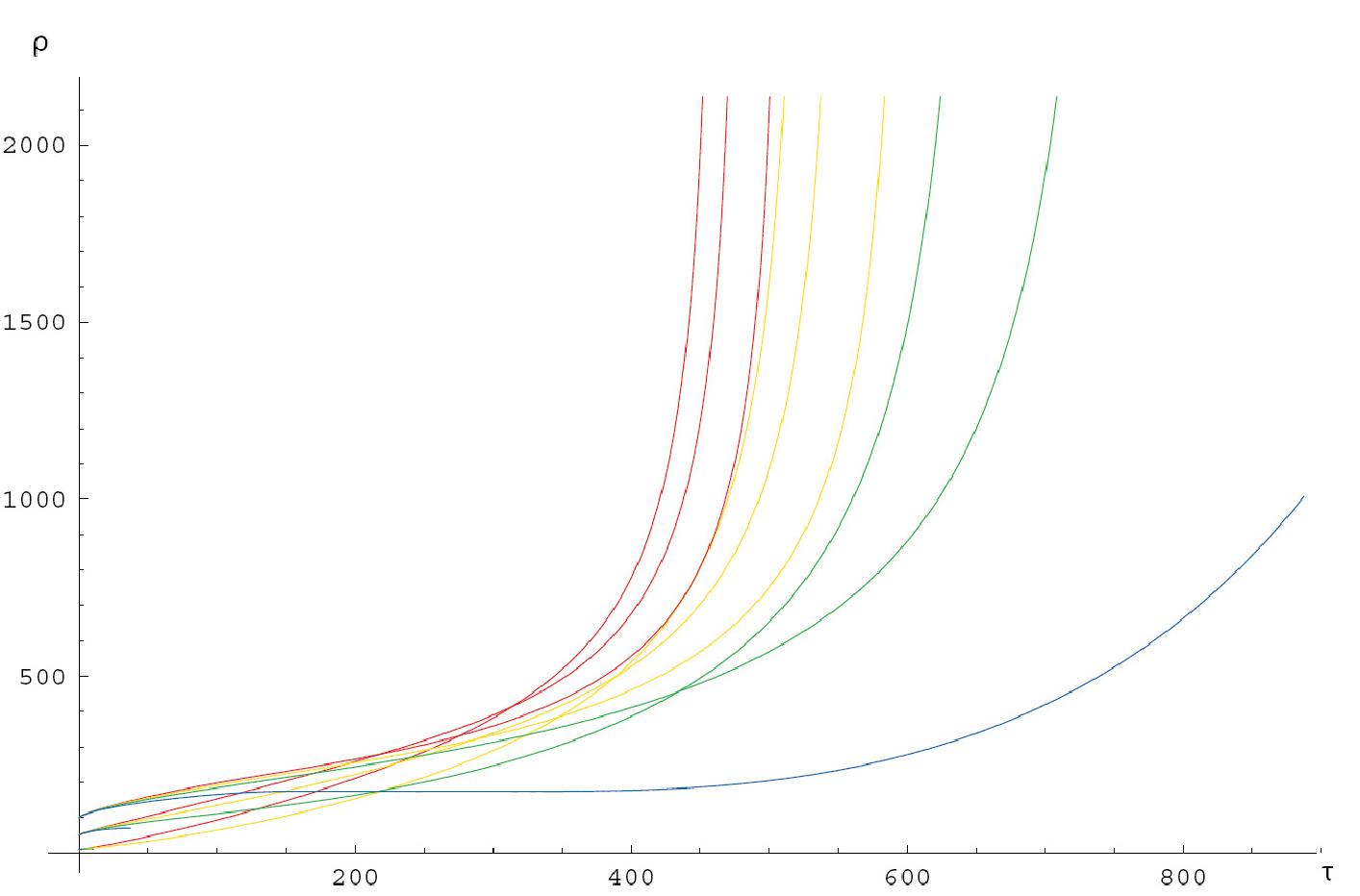}
\end{minipage}\\[-10pt]
\hspace{-1.2cm}\begin{minipage}[t]{.4\textwidth}
\caption{
Time evolution of the bubble in the outside coordinates, $x[t]$, for several initial sizes,
$x_{init}=10,50,100$. The background contains
homogeneous curvature ($R=120$) and matter ($A=10^{-4}$).
Colors correspond to different equations of state; $w=1/3$ red, $w=0$ yellow, $w=-1/3$ green, $w=-1$ blue.
}\label{figure_x_fixed_120}
\end{minipage}%
\hspace{1.2cm} \begin{minipage}[t]{.4\textwidth}
\caption{
Time evolution of the bubble in the bubble coordinates, $\rho[t]$, for the same background conditions as in
Figure \ref{figure_x_fixed_120}.
Colors correspond to different equations of state; $w=1/3$ red, $w=0$ yellow, $w=-1/3$ green, $w=-1$ blue.
} \label{figure_rho_fixed_120}
\end{minipage}%
\end{figure}

Next we turn to the case with more curvature, but still such that $R > R_{cr}$.
From Figure \ref{figure_x_fixed_120} we see that in most cases bubbles again grow, but now
they asymptote to the upper bound, $x=R$. 

The corresponding bubble evolution in the coordinates on the bubble is shown in
Figure \ref{figure_rho_fixed_120}. The overall effect of more curvature is a further
slowdown in the time evolution of the bubbles.

{\em In summary}, in this section we have considered a closed FRW background that contains dust as well as a cosmological constant but that will eventually asymptote to a de Sitter space. In this case it is no longer consistent to have vacuum shells separating de interior and exterior region \footnote{The point of this section is to consider bubble creation for times sufficiently small for the dust density to make a difference. For long times the dust density will redshift and this background will become indistinguishable from the previous one.}. We have studied the motion  for different values of $w$. In shell coordinates, for $w=-1$, we have found that some bubbles do eventually become subcritical or stabilize and stop contracting.

\subsubsection {$R < R_{cr}$}

\begin{figure}[htp]
\centering
\includegraphics [width = 10 cm]{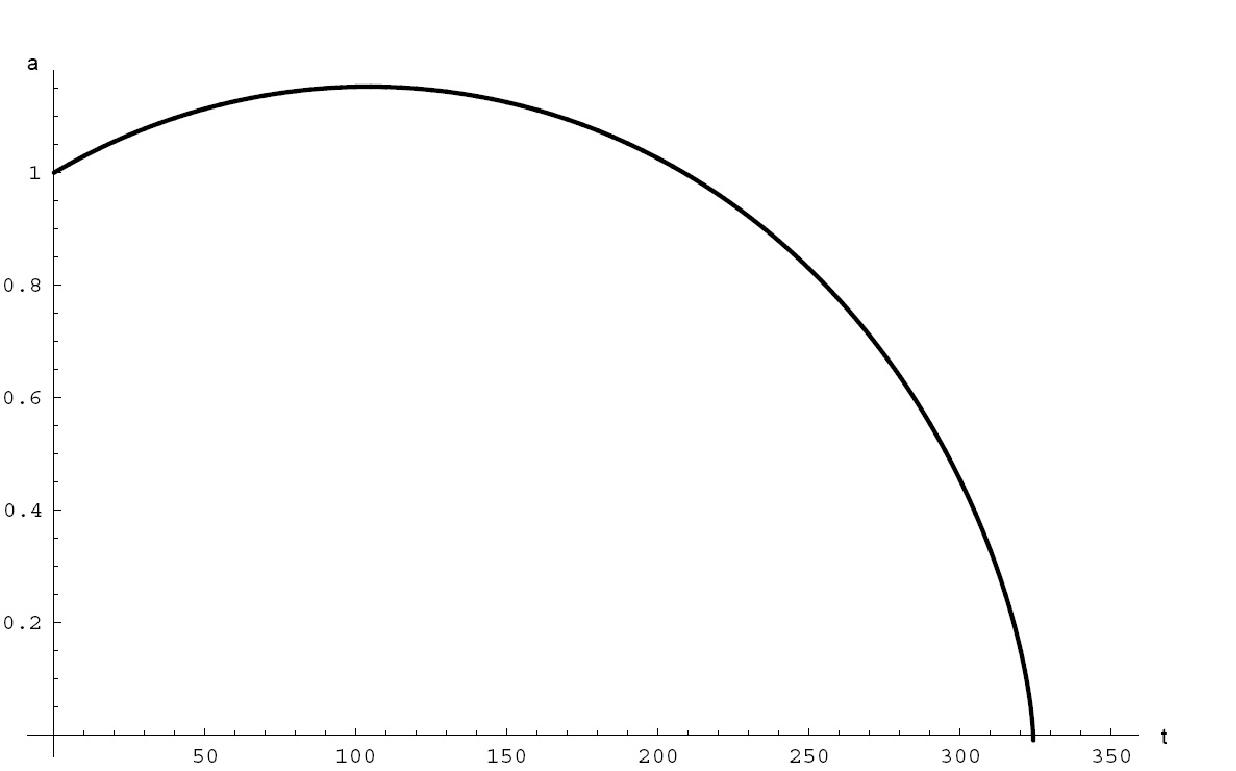}
\caption{Time evolution of the scale factor $a[t]$ for $R=100$. ($\Lambda =3 \times 10^{-5}, A=10^{-4}$.)}
\label{figure_a_crunch_100}
\end{figure}

For curvature greater than the critical value (i.e. $R < R_{cr}$) the space will
eventually collapse. To investigate this case we choose $R=100$. 
The evolution of the scale factor is shown in Figure \ref{figure_a_crunch_100}.

\begin{figure}
\centering
\hspace{-0.8cm}\begin{minipage}[b]{.5\textwidth}
\centering
\includegraphics[width=6cm]{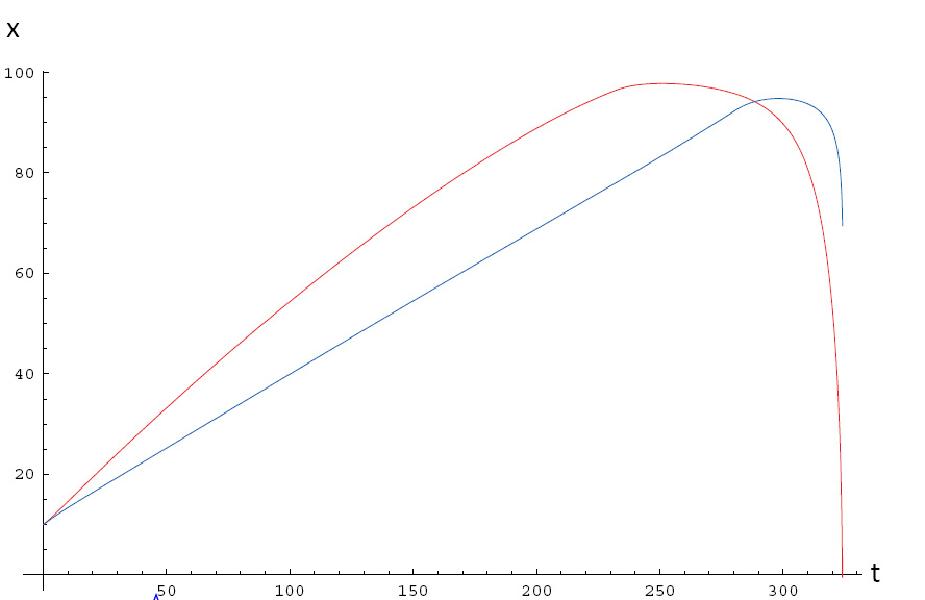}
\end{minipage}%
\begin{minipage}[b]{.55\textwidth}
\centering
\includegraphics[width=6cm]{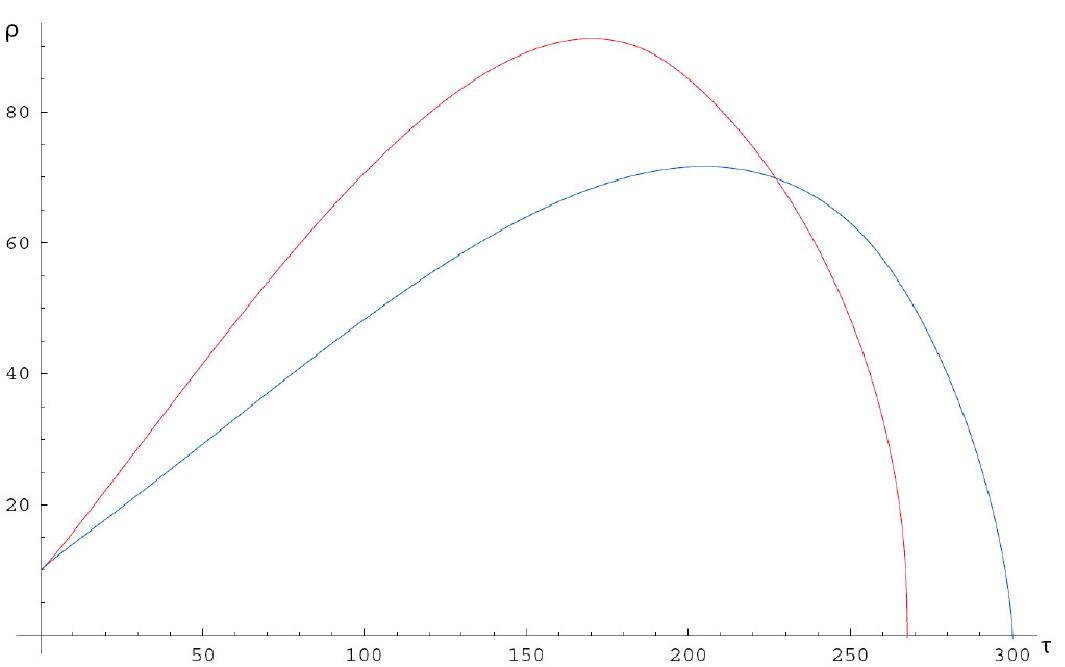}
\end{minipage}\\[-10pt]
\hspace{-1.4cm}\begin{minipage}[t]{.4\textwidth}
\caption{
Time evolution of the bubble in the outside coordinates, $x[t]$, for crunching background (Figure \ref{figure_a_crunch_100}).
$x_{init}=10$. $R=100$. Colors correspond to different equations of state: $w=1/3$ red, $w=0$ blue.
}\label{figure_x_crunch_cont_100}
\end{minipage}%
\hspace{1.8cm} \begin{minipage}[t]{.4\textwidth}
\caption{
Time evolution of the bubble in the bubble coordinates, $\rho[t]$, for crunching background
(Figure \ref{figure_a_crunch_100}). $x_{init}=10$. $R=100$. Colors correspond to different equations of state; $w=1/3$ red, $w=0$ blue.
} \label{figure_rho_crunch_cont_100}
\end{minipage}%
\end{figure}

In the outside coordinates, the time evolution of most bubbles will once again
lead to the upper bound, $x=R$ (Figure \ref{figure_x_crunch_cont_100}). 
Bubbles with $w=-1$ contract and hit the lower bound $x=\frac{1}{aB}$, 
which stops their further evolution.

Figure \ref{figure_rho_crunch_cont_100} reveals corresponding behavior in the
coordinates on the bubble. Bubbles which do not hit the bound $x=\frac{1}{aB}$
eventually collapse along with the collapse of the space itself.

{\em In summary}, in this section we have considered a closed FRW background that contains dust and a cosmological constant and that will eventually contract. As in the previous case it is not longer consistent to have vacuum shells separating the inside and outside region. Depending on the initial size, bubbles with smaller value of $w$ become subcritical and disappear while the others continue to grow and eventually collapse along with the space itself. 

\section{\bf Bubble Expansion in Inhomogeneous Backgrounds}

\subsection{Generating a Curvature Profile}

In order to generate an inhomogeneous curvature profile, we choose:

\eqn{R(r)}{R(r)=(\alpha r + \beta R_{cr})\left( 1\pm \frac{1}{\gamma + (\delta -r)^2} \right)}
in the definition of the LTB metric. 
Choosing $\beta >1$ will allow us to generate a sharp drop in the function $R(r)$, for
the cases where the minus sign is chosen. The position and the width of the extremum
will be regulated by the parameter $\delta$, and the depth/height by $\gamma$. 

Furthermore, we want to make sure to satisfy the weak energy condition, namely
that the matter density stays positive:

\eqn{d(t,r)}{d(t,r)=\frac{3A}{a^2(t,r) (a(t,r) + r a'(t,r))}}
Since $A$ is positive definite, this gives us a condition:

$$a(t,r) + r a'(t,r)>0$$
This means that $a(r)$ should nowhere fall faster than $1/r$.

It should be noted that (\ref{R(r)}) incorporates an approximation to a 
delta 
function in the curvature profile. Since a constant value of $R(r)$ can be 
re-absorbed by a variable redefinition in the metric, we are interested 
in the physics associated with a spatially varying $R(r)$. In regions 
of spacetime where $R(r)$ is roughly constant, we expect that the 
evolution 
of the bubble would be as in the FRW case\footnote{A crucial ingredient 
in coming to this conclusion is the fact that the LTB scale factor can 
be solved as an {\em ordinary} differential equation in time $t$, point by 
point in the spatial grid. See Section 
2.}. In the next 
sub-section, two representative examples which capture the essential 
features of these gradient effects are presented.

\subsection{Examples}

\begin{figure}[htp]
\centering
\hspace{-0.8cm}\begin{minipage}[b]{.5\textwidth}
\centering
\includegraphics[width=6cm]{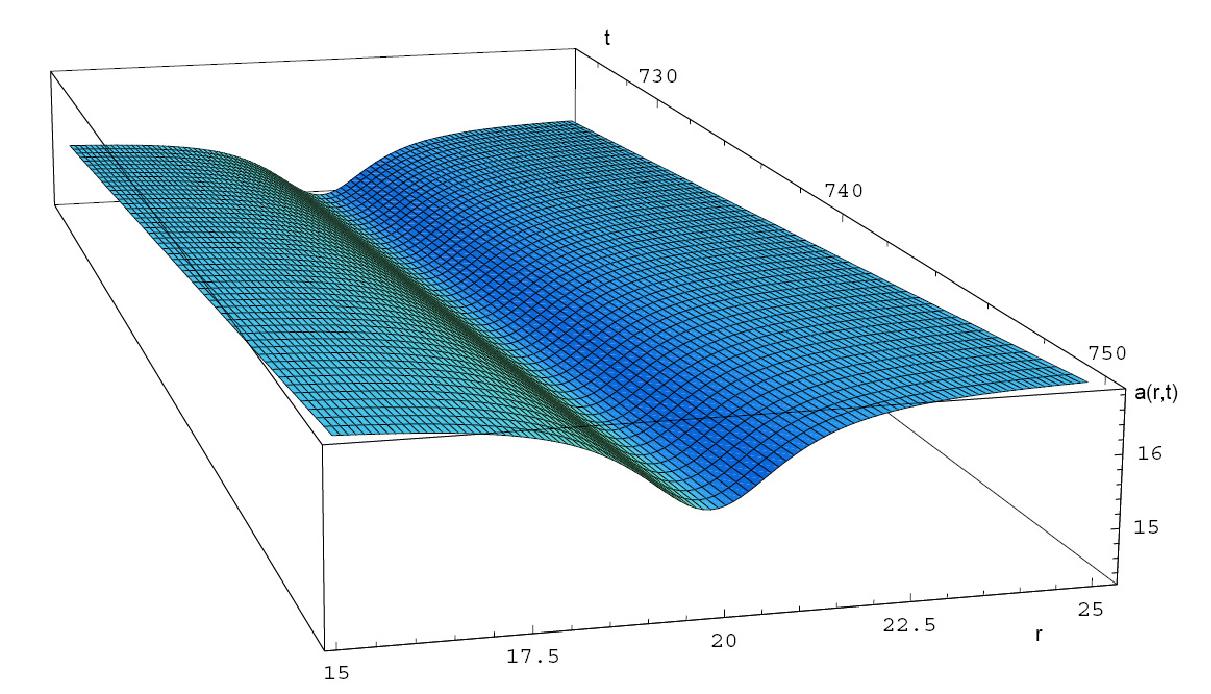}
\end{minipage}%
\begin{minipage}[b]{.55\textwidth}
\centering
\includegraphics[width=6cm]{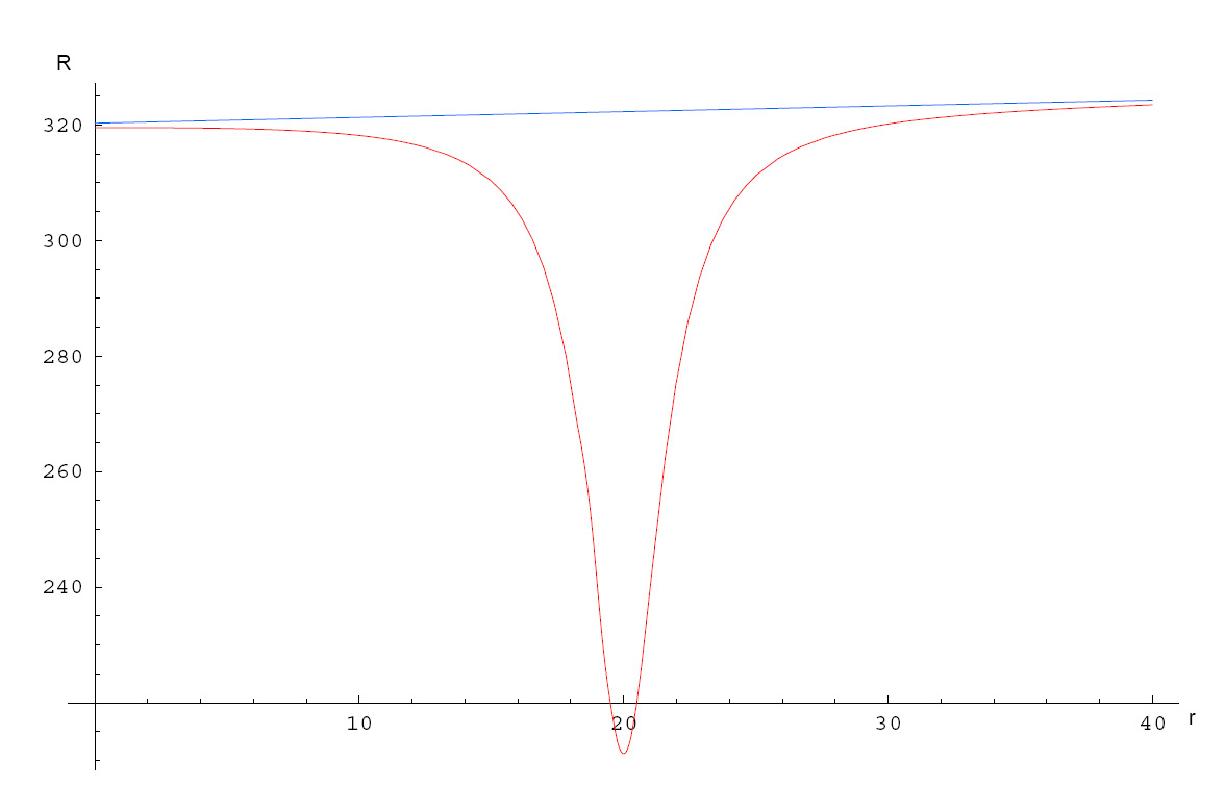}
\end{minipage}\\[-10pt]
\hspace{-1cm}\begin{minipage}[t]{.4\textwidth}
\caption{
The position dependent scale factor $a(r,t)$ of the the LTB spacetime, for 
the curvature profile depicted in red in Figure \ref{figure_R_prof_m}.
}\label{a_prof}
\end{minipage}%
\hspace{2cm} \begin{minipage}[t]{.4\textwidth}
\caption{
$R(r)$ works as an inhomogeneity profile. Two choices are shown 
here, the red curve results in an inhomogeneous spacetime.
} \label{figure_R_prof_m}
\end{minipage}%
\end{figure}

\begin{figure}[htp]
\centering
\hspace{-0.8cm}\begin{minipage}[b]{.5\textwidth}
\centering
\includegraphics[width=6cm]{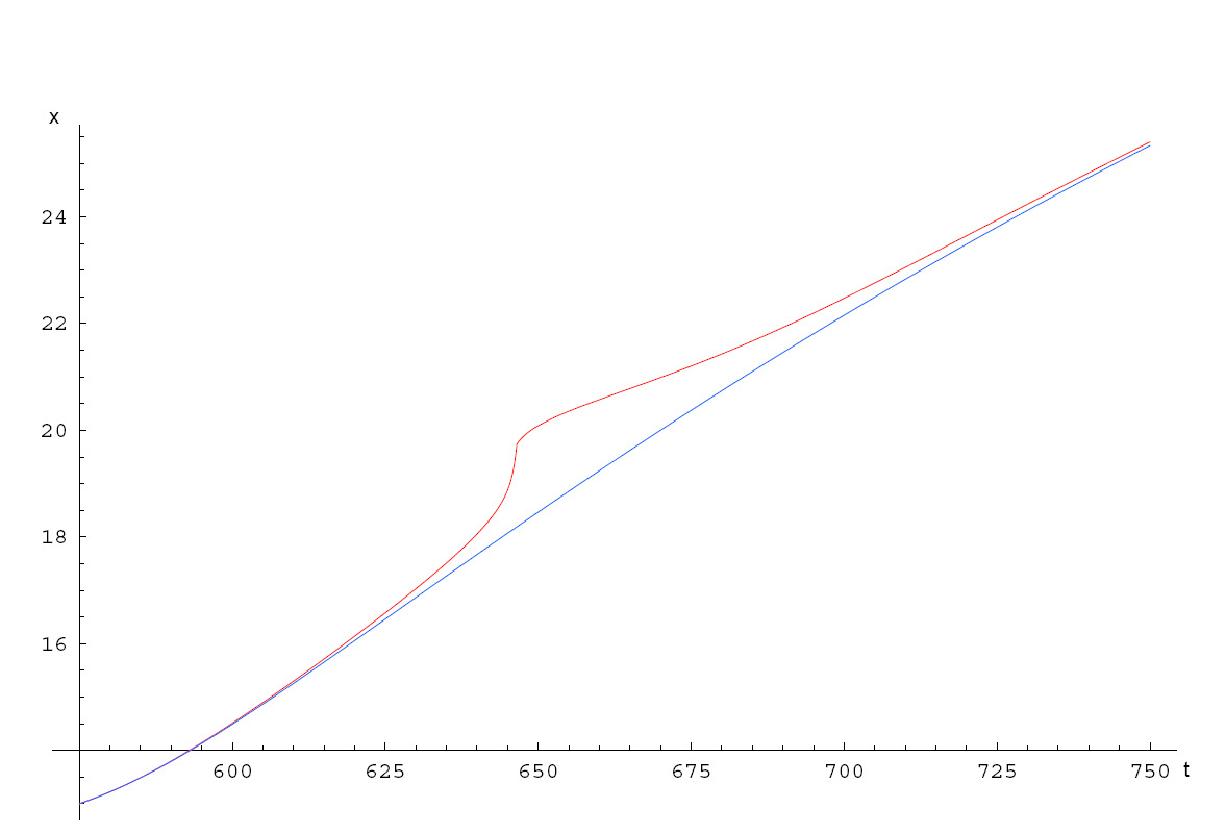}
\end{minipage}%
\begin{minipage}[b]{.55\textwidth}
\centering
\includegraphics[width=6cm]{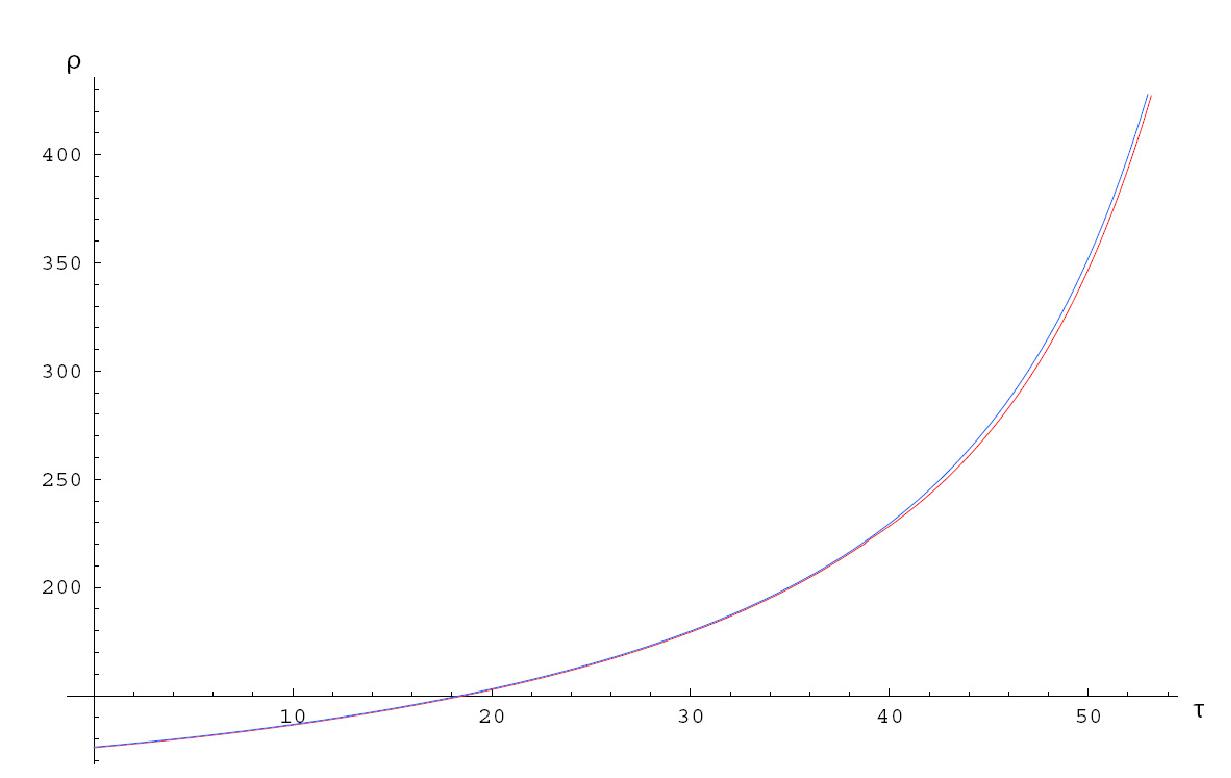}
\end{minipage}\\[-10pt]
\hspace{-1cm}\begin{minipage}[t]{.4\textwidth}
\caption{
Evolution of bubbles in outside coordinates, $x(t)$, for the two curvature 
profiles shown in Figure \ref{figure_R_prof_m}.
}\label{figure_x_prof_m}
\end{minipage}%
\hspace{2.0cm} \begin{minipage}[t]{.4\textwidth}
\caption{
Evolution of bubbles in bubble coordinates, $\rho(\tau)$, for the two 
curvature profiles shown in Figure \ref{figure_R_prof_m}.
} \label{figure_rho_prof_m}
\end{minipage}%
\end{figure}

As an example, we first choose the minus sign in $R(r)$, with $\alpha =0.1$, $\beta = 3$ and $\delta = 20$.
In order for $R$ to stay above $R_{cr}$, in this case we need $\gamma > 1.5$.
In addition, numerical simulations show that to satisfy the weak energy condition,
$\gamma \geq 2.9$, and since we'd like to investigate the sharpest possible profile,
we choose $\gamma = 2.9$.

This choice of the function $R(r)$ generates a profile in the evolution
of the scale factor $a(r,t)$, as shown in Figure \ref{a_prof}. We will compare evolution
of a bubble in such background with evolution in a curvature background 
without a fluctuation in the profile. The two choices of $R(r)$ are shown 
in Figure \ref{figure_R_prof_m}.
The resulting bubble evolutions are shown in Figures \ref{figure_x_prof_m} and \ref{figure_rho_prof_m}.
We see that even though in the outside coordinates bubble evolution is greatly affected by the curvature profile,
such effect is absent for the bubble evolution in the coordinates on the bubble.

\begin{figure}[htp]
\centering
\hspace{-0.8cm}\begin{minipage}[b]{.5\textwidth}
\centering
\includegraphics[width=6cm]{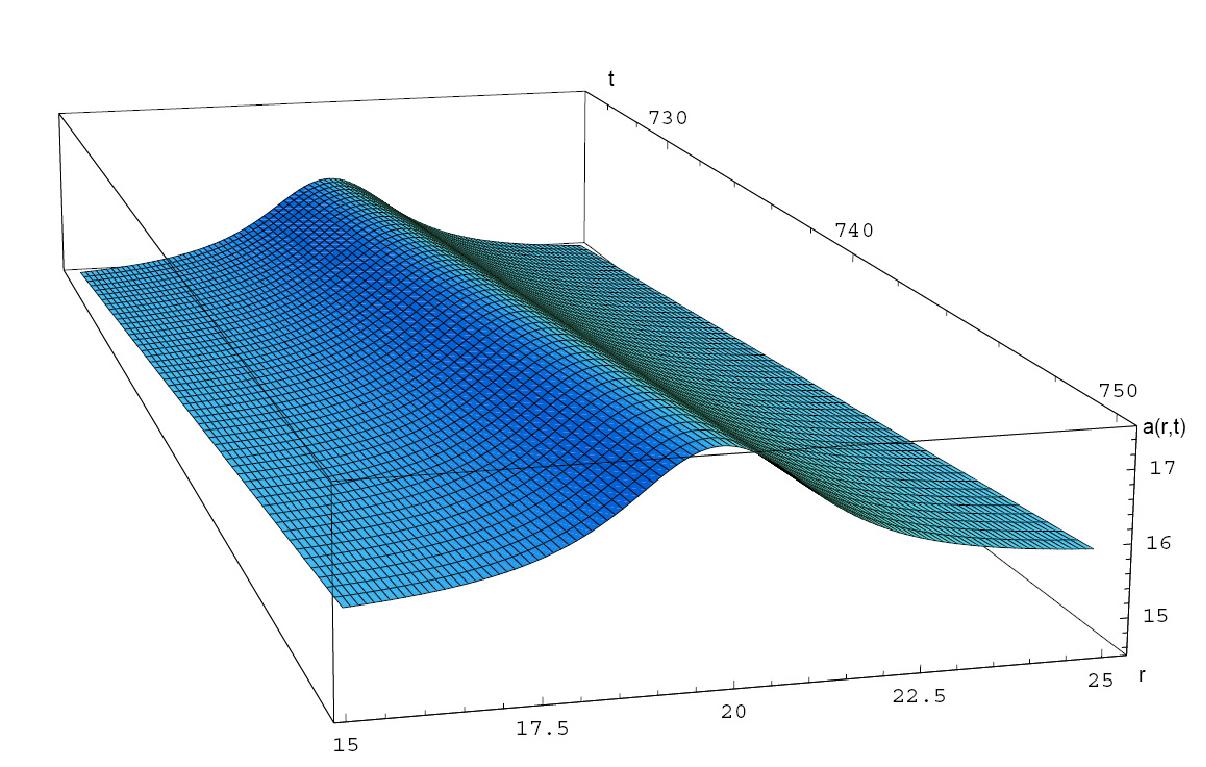}
\end{minipage}%
\begin{minipage}[b]{.55\textwidth}
\centering
\includegraphics[width=6cm]{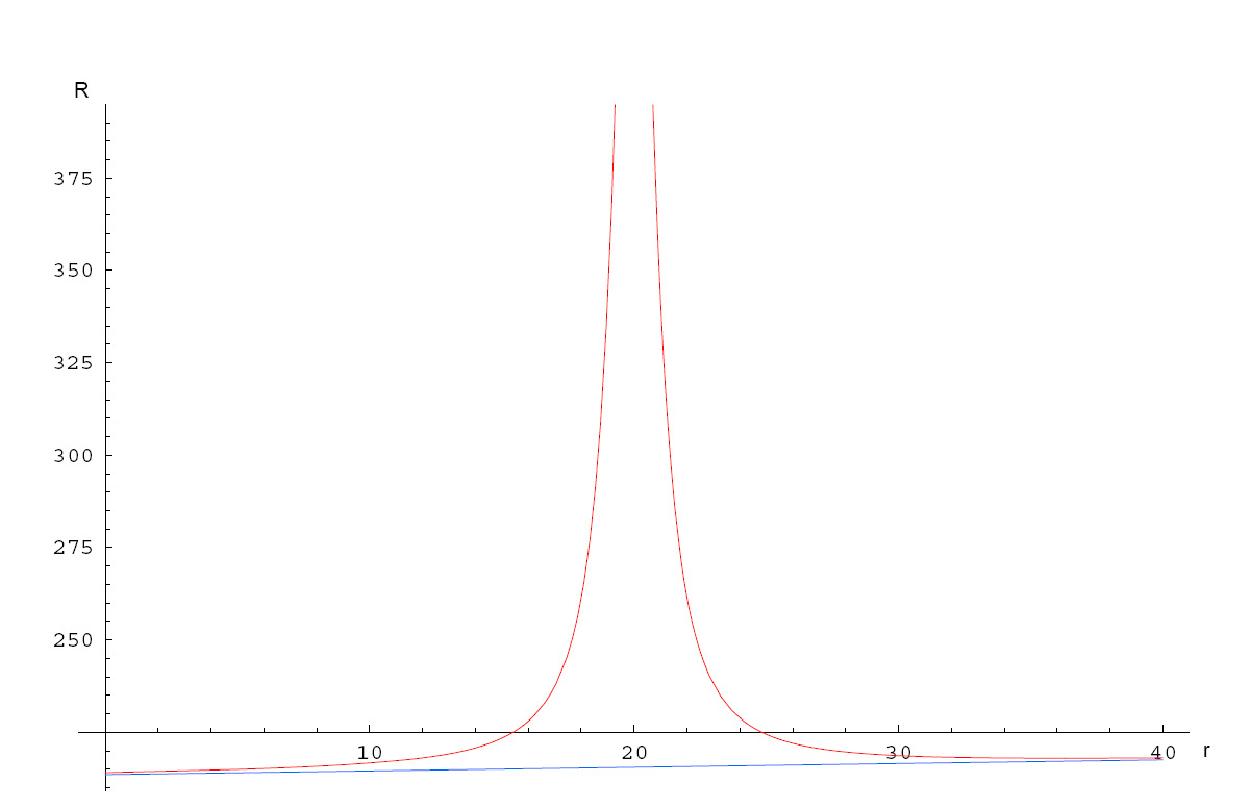}
\end{minipage}\\[-10pt]
\hspace{-1cm}\begin{minipage}[t]{.4\textwidth}
\caption{
$a(r,t)$ for curvature profile depicted in red in Figure 
\ref{figure_R_prof_m_up}
}\label{a_prof_up}
\end{minipage}%
\hspace{2cm} \begin{minipage}[t]{.4\textwidth}
\caption{
Second example of an inhomogeneous background: Two choices of $R(r)$ are 
shown.
} \label{figure_R_prof_m_up}
\end{minipage}%
\end{figure}

\begin{figure}[htp]
\centering
\hspace{-0.8cm}\begin{minipage}[b]{.5\textwidth}
\centering
\includegraphics[width=6cm]{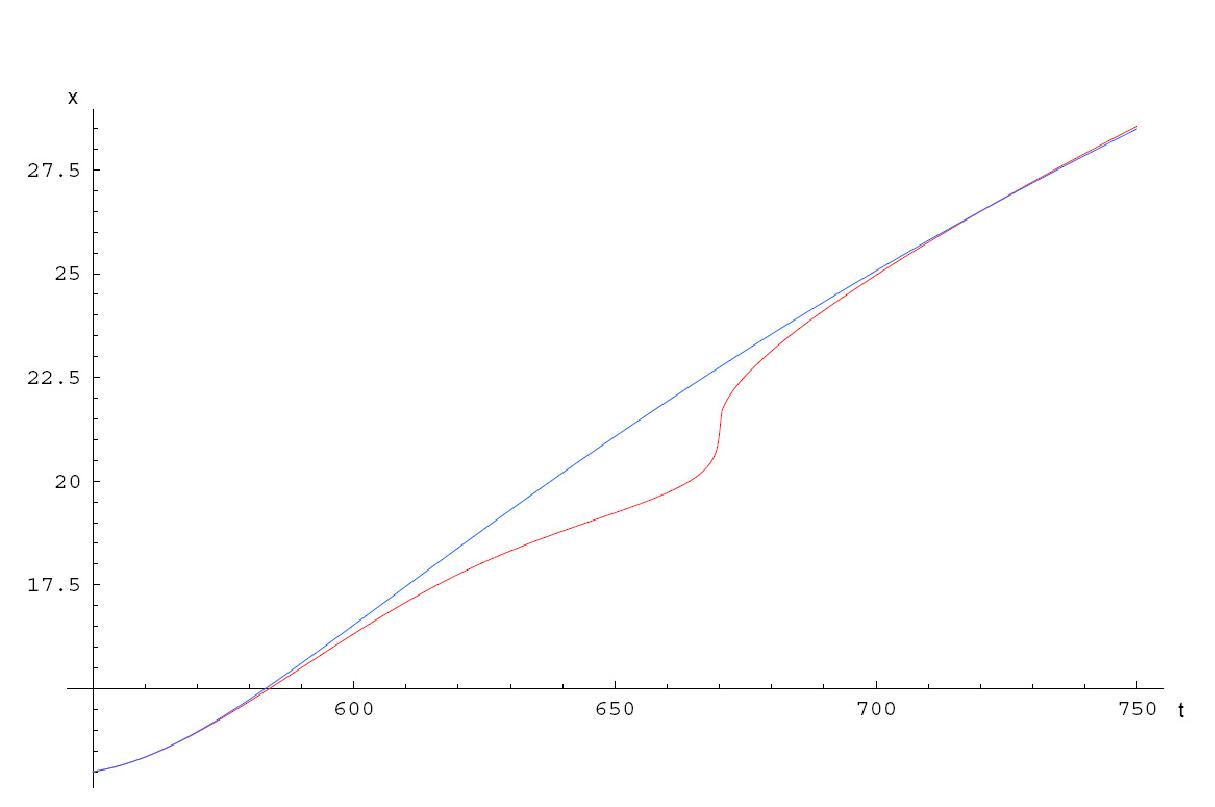}
\end{minipage}%
\begin{minipage}[b]{.55\textwidth}
\centering
\includegraphics[width=6cm]{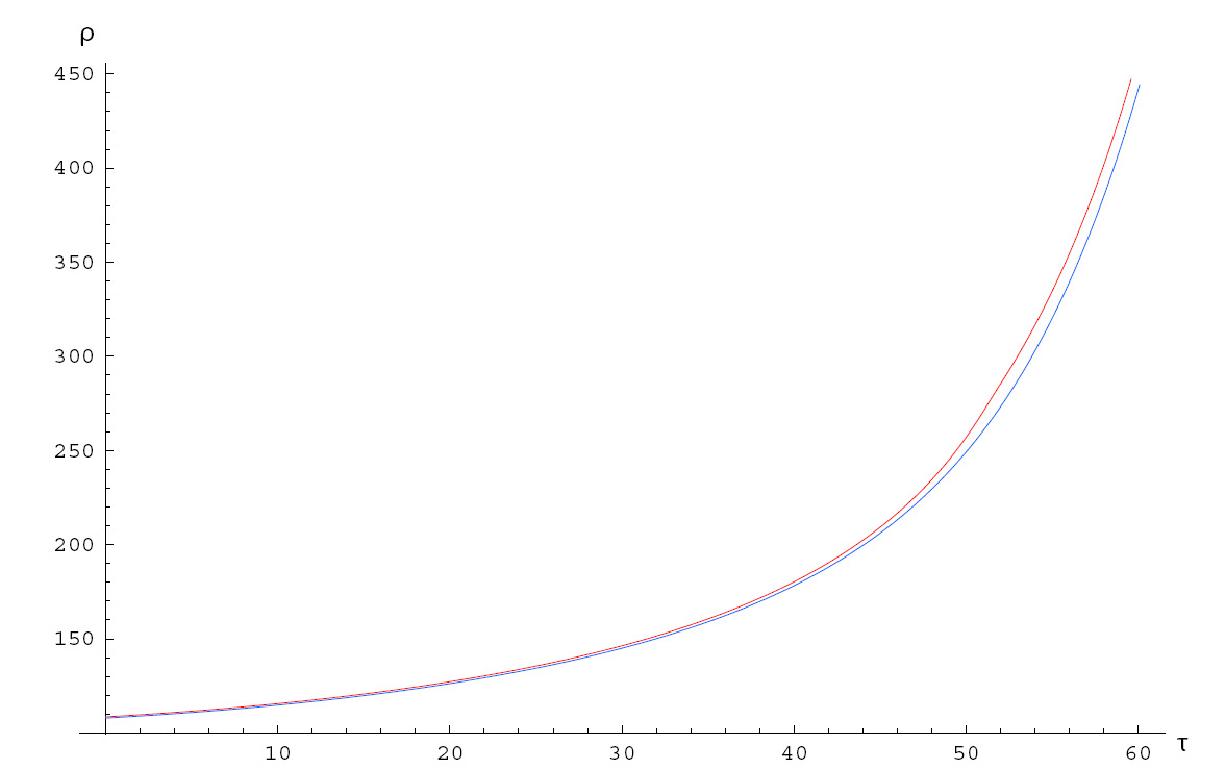}
\end{minipage}\\[-10pt]
\hspace{-1cm}\begin{minipage}[t]{.4\textwidth}
\caption{
Evolution of bubbles in the outside coordinates, $x(t)$, for the two 
curvature 
profiles shown in Figure \ref{figure_R_prof_m_up}.
}\label{figure_x_prof_m_up}
\end{minipage}%
\hspace{2cm} \begin{minipage}[t]{.4\textwidth}
\caption{
Evolution of bubbles in the shell coordinates, $\rho(\tau)$, for the two 
curvature profiles shown in Figure \ref{figure_R_prof_m_up}.
} \label{figure_rho_prof_m_up}
\end{minipage}%
\end{figure}

For comparison, we will also choose $R(r)$ with the plus sign, and
$\alpha =0.1$, $\beta = 2$,$\gamma=0.7$ and $\delta = 20$, depicted in  Figure \ref{figure_R_prof_m_up}.
This results in a profile for $a(r,t)$ shown in Figure \ref{a_prof_up}. 
The corresponding bubble evolution
is shown in Figures \ref{figure_x_prof_m_up} and \ref{figure_rho_prof_m_up}.
Again, we find that the fluctuation does not qualitatively alter the 
bubble evolution.

\section{\bf Conclusions}
We have examined the evolution of bubbles of true vacuum in several
backgrounds (that always include a cosmological constant) to assess the
effect of curvature on the propagation. 
The bubble evolution changes qualitatively when looked at in terms of the 
outside coordinates, but when studied in terms of the more physically 
relevant coordinates on the shell or its interior, the local curvature is 
largely irrelevant. In the presence of 
matter, we observe that 
some bubbles, with an equation of state close to $w=-1$, disappear. Those 
that survive do behave qualitatively the same. 

The explanation for this insensitivity to inhomogeneities is to
be found in the weak energy condition: our background spacetime is
consistent only when (\ref{d(t,r)}) is positive, which restricts the
sharpness of the inhomogeneity profiles that are allowed. This could of
course be a restriction arising from the ans\"atz for the background that
we are working with. But it should be mentioned that in the work of  Wald
\cite{Wald:1983ky} that was mentioned in the introduction, the
energy conditions played a crucial role in the proof of the cosmic no
hair theorem for anisotropic (but homogeneous) cosmologies.

We view this work as an attempt to test the requirements to arrive at a 
universe like ours without making the simplifying assumptions of starting 
with an FRW metric.  There are many directions in which the work here can 
be extended. We have
examined only a few inhomogeneity profiles here, a more thorough
understanding of the evolution curves in more generic situations is
certainly of interest. For all the various initial conditions and 
collapse/inhomogeneity profiles that are allowed, the question of bubble 
evolution should be tractable with our 
machinery\footnote{Though such a scan would admittedly be much more 
numerically challenging than the few cases we have undertaken here.}. 
If we think of our setup as a toy model for inflation, it will 
definitely be of interest to study the {\em genericity} of inflation in 
the space of possible inhomogeneity profiles.  
Another important direction to follow is to evaluate the tunneling 
amplitudes for inhomogeneous ambient spaces. This is a notoriously 
difficult problem on which any enlightenment will be welcome. 

Another problem that is of interest in this context is whether there are
signatures left on the new (inside) Universe due to the propagation of the
bubble through the inhomogeneous ambience. The bubble can be thought of
as a Casimir cavity with a moving wall and the cavity radiation (the
analogue of the CMB in this context) will carry imprints of the external
inhomogeneities. Because of the spherical symmetry, this is effectively a
two-dimensional QFT problem, with moving mirror boundary conditions
\cite{Birrell}.

\section{Acknowledgments}

CK would like to thank Frank Ferrari for useful discussions, and the 
audience at the Joint ULB-VUB-KUL-Solvay seminar for feedback on a talk 
based on this paper. The work of CK is supported in 
part by IISN - Belgium (convention 4.4505.86), by the Belgian National 
Lottery, by the European Commission FP6 RTN programme MRTN-CT-2004-005104 
in which CK is associated with V. U. Brussel, and by the Belgian Federal 
Science Policy Office through the Interuniversity Attraction Pole P5/27. 

The work of W. Fischler, S. Paban and M. \v{Z}ani\'{c} is
partially supported by the National Science Foundation under Grant No. PHY-0455649.
 
\section{\bf Appendix}

In this appendix, we will derive the coupled differential equations that
form the starting point for our numerical shell evolution plots. These
equations are a direct consequence of Israel's junctions
conditions\footnote{A pedagogical
derivation of the junction conditions based on a distributional approach
to Einstein's equations
can be found in \cite{Poisson(2004)}.
}, which relate the discontinuity in the extrinsic curvature {\em across}
the shell surface to the energy-momentum layer {\em on} the shell. The
general expressions for these equations for the case
of spherically symmetric junctions
can be found in,
e.g., \cite{Berezin:1987bc}. Here we discuss a few of the ingredients
that go into our specific problem by doing some illustrative computations
{\em ab initio}. The aim is to stress some issues like choices of signs
and coordinates which are often not adequately addressed in the
litearture.

We will work with the mostly minus $\{+,-,-,- \}$ metric,
and the signs on the energy-momentum tensor for the ideal fluid are fixed
by
\eqn{energy}{T_{\alpha}^{\ \beta}=(\epsilon+p)u_{\alpha}
u^{\beta}-p\delta_{\alpha}^{\ \beta}.}
Einstein's equation takes the form $G_{\alpha \beta}=T_{\alpha
\beta}$, in natural units ($8\pi G=c=1$). Later, when we work with
quantities defined on the shell, we will have analogous sign conventions
for them as well. Throughout, Latin indices denote
3-dimensional objects defined on the
shell hypersurface, Greek indices stand for 4-dimensional
quantities, and semi-colon is shorthand for the covariant derivative. We
will also need
$e^\alpha_a\equiv \frac{\partial x^\alpha}{\partial y^b}$, which  are
projectors
(pull-backs) that can be used to project a 4-quantity on to the 3-surface.

The first of Israel's two conditions says that the metric induced on the
shell from the bulk 4-metrics on either side should match, and be equal
to the 3-metric on the shell.
The assumption of spherical
symmetry restricts the form of
the intrinsic metric on the shell to the form,
\eqn{shellmetric}{{ds_3}^2=d\tau^2-{\rho(\tau)}^2d\Omega_2,}
where $\tau$ is the only independent coordinate on the
shell, which we take to be the shell proper time. So looked at from the
outside, on the shell, we can parameterize the
coordinates as
$r=x(\tau)$ and $t=t(\tau)$, and from the inside,
$T=T(\tau)$ and $z=z(\tau)$. Since the metric from either side on the 
shell
should agree with the 3-metric on the shell,
we get two conditions from the inside
(open-FRW),
\eqn{mink-match1}{b(T)z=\rho(\tau), \ \Big(\frac{dT}{d\tau}\Big)^2=
1+\frac{b(T)^2}{1+z^2}\Big(\frac{dz}{d\tau}\Big)^2,}
and two from the outside (LTB),
\eqn{peeb-match1}{a\big(t,x\big)x=\rho(\tau), \
\Big(\frac{dt}{d\tau}\Big)^2=1+\frac{{\big(a(t,x)x\big),_{x}}^2}{1-x^2/R(x)^2}\Big(\frac{dx}{d\tau}\Big)^2.}
All the variables in these equations are thought of as functions of 
$\tau$.

Now we turn to the second junction conditions, which determine the
dynamics of the shell. These are expressed in
terms of the extrinsic curvature:
\eqn{defineK}{K_{ab}=n_{\alpha;\beta} \ e^\alpha_a \ e^\beta_b.}
Here $n_\alpha$ is the outward normal to the
surface under question (since we are working with a closed shell, there
is no ambiguity in making this choice). There
are many different choices and sign
conventions for the extrinsic curvature in the literature, we have made
the above definitions so that the extrinsic curvature of a
2-sphere is positive. In particular, this differs from
\cite{Berezin:1987bc},
but is in agreement with most of the other references/reviews on the 
subject \cite{Poisson(2004), Lake2, 
LagunaCastillo:1986je, Barrabes:1991ng}. With 
these prescriptions, the
second
junction condition takes the form
\eqn{irael-g}{[K_{ab}]-h_{ab}[K]=S_{ab},}
with $h_{ab}$ denoting the shell metric, which in our case is
(\ref{shellmetric}). The 3-tensor $S_{ab}$ stands for the surface
energy-momentum tensor, and we will assume it to have a perfect-fluid form
analogous to (\ref{energy}). Square brackets stand for discontinuities
across the shell: $[K]=K_{out}-K_{in}$.

It can be shown \cite{Berezin:1987bc}, that the junctions conditions imply
a conservation
law of
the form
\eqn{conserve}{{S_a^{\ b}}_{; b}+[e^\alpha_a T^{\beta}_{\alpha}
n_\beta]=0,}
where $T^{\beta}_{\alpha(out/in)}$ is the bulk energy-momentum tensor on 
either
side.
The advantage of this equation is that it is first order, and one can
exchange\footnote{This is analogous to the fact that the covariant
energy conservation equation is an integrability condition for Einstein's
equations.} one of
the second order equations arising from the
junction conditions with (\ref{conserve}). This is indeed what we will
do.


Because of spherical symmetry and the form of
the
three metric, the extrinsic curvature $K_a^{\ b}$ has
independent components
$K_\tau^{\ \tau}$
and $K_\theta^{\ \theta}=K_\phi^{\ \phi}$, while the surface energy tensor
$S_a^{\ b}$
contains $S_\tau^{\ \tau}\equiv \sigma$ and
$S_\theta^{\ \theta}=S_\phi^{\ \phi}\equiv -P$. So the
independent relations that
arise from the second junction conditions are
\begin{eqnarray}\label{sigmaequation}
-\frac{\sigma}{2}&=&[K_\theta^\theta],\\
P&=&[K_\tau^\tau]+[K_\theta^\theta]\label{pequation}.
\end{eqnarray}
Both $\sigma$ and $P$ are purely functions of
$\tau$ by spherical symmetry. The conservation law (\ref{conserve}) takes
the form,
\eqn{Ton}{\frac{d\sigma}{d\tau}+\frac{2}{\rho}\frac{d\rho}{d\tau}(\sigma+P)
+[T_\tau^n]=0, }
where $[T_\tau^n] \equiv [e^\alpha_\tau T^{\beta}_{\alpha}
n_\beta]$. The evolution of the shell is completely determined by
(\ref{sigmaequation}) and (\ref{Ton}), so our task then is to write down
these differential equations explicitly so that we can proceed with the
numerics.

We have different
sets of coordinates in each of the three regions, but not all
of these coordinates can be explicitly written in terms of the others. So 
we will 
write
the matching conditions in terms of the LTB
coordinates. The LTB metric is known only numerically
(see (\ref{EOM}), (\ref{EOMB})), so there
is no hope of writing it in terms of the other coordinates, thereby 
making this
choice inevitable.
Since the evolution is best
understood in the shell coordinates, once we have the evolution curves in
LTB coordinates, we will translate
them numerically to $\rho(\tau)$.

We start with (\ref{sigmaequation}). 
To calculate the extrinsic curvatures, we need the normal vectors in
the corresponding coordinates. We will start with the LTB side where
the coordinates are $x^\alpha=(t,x,\theta,\phi)$. The projectors are:
\begin{eqnarray}
u^\alpha\equiv
e^\alpha_\tau=&&\hspace{-0.1in}\Big(\frac{dt}{d\tau},\frac{dx}{d\tau},0,0\Big),
\nonumber \\
e^\alpha_\theta=\big(0,0,1,0\big), && \ \
e^\alpha_\phi=\big(0,0,0,1\big). \nonumber
\end{eqnarray}
Since $u^{\alpha}$ is the 4-velocity of the bubble, the normal $n_{\beta}$
will be determined (upto a sign) by the two conditions $u^\alpha
n_\alpha=0$ and $n^\alpha n_\alpha=-1$, where raising and lowering are
done with the LTB metric. The second condition arises because our
shell is timelike: a timelike shell is defined by a spacelike
normal. A simple calculation yields,
\eqn{peeble-normal}{{n_\alpha}=\gamma_{out}\left(\frac{-(ax),_{x}{\dot
x}}{\sqrt{(1-x^2/R^2)}}
,\frac{(ax),_{x}{\dot t}}
{\sqrt{(1-x^2/R^2)}}
,0,0\right)}
where the dots are with respect to $\tau$. 
The
choice of sign for the normal is encoded in $\gamma_{out}$, and is fixed
by the condition that the normal should point from the inside to the
outside. For an expanding shell, this
means that $\gamma_{out}=+1$. For a collapsing shell $\gamma_{out}=
-1$. 

Using these, the $K^\theta_\theta$ on the LTB side turns out to be,
\eqn{Ktheta+}{ K^{\theta}_{\theta (out)}
=h^{\theta\theta}n_{\theta;\theta}=\frac{1}{\rho^2}\Big(n_{\theta,\theta}-
\Gamma^\alpha_{\theta\theta}n_{\alpha}\Big)
=\gamma_{out}\Big(\frac{(ax),_t(ax),_x {\dot x}+(1-x^2/R^2){\dot 
t}}{\rho
\sqrt{(1-x^2/R^2)}}\Big).}
We have used (\ref{peeb-match1}) to do some of the simplifications, 
and $(ax),_t$ stands for $(x \partial_t a)$ because at this stage $x$ and 
$t$ are unrelated variables: when we eliminate $\tau$ to write $x$ as a 
function of $t$, this will no longer be the case. 
Repeating the above calculation for 
$K_\theta^\theta$ on the
inside (open-FRW), with coordinates $x^{\alpha}=(T, z, \theta, \phi)$, we
have
\eqn{mink-K}{K_{\theta (in)}^\theta=\gamma_{in}\Big(\frac{z b
\frac{db}{dT}\dot z
+(1+z^2) \dot T
}{\rho\sqrt{1+z^2}}\Big).}
Again, $\gamma_{in}=+1$ when the radius of the inside region is 
increasing; otherwise, $\gamma_{in}=-1$. 
The junction condition becomes,
\eqn{mixed}{\gamma_{out}\Big(\frac{(ax),_t(ax),_x {\dot
x}+(1-x^2/R^2){\dot
t}}{\rho
\sqrt{(1-x^2/R^2)}}
\Big)-
\gamma_{in}\Big(\frac{z b \frac{db}{dT}\dot z
+(1+z^2) \dot T
}{\rho\sqrt{1+z^2}}\Big)=-\frac{\sigma}{2}.}
By using (\ref{peeb-match1}), it is possible to rewrite this
equation in a more standard form as:
\eqn{standard-junc}{\gamma_{out}\sqrt{{\dot
\rho}^2-\Delta_{out}}-\gamma_{in}\sqrt{{\dot
\rho}^2-\Delta_{in}}=-\frac{\sigma\rho}{2},}
where $\Delta_{out}=-(1-x^2/R^2)+{(ax),_t}^2$, and
$\Delta_{in}=-(1+z^2)+z^2(db/dT)^2$. The quickest way to demonstrate this 
is to start with the final expressions. For instance, for the LTB piece, 
we can expand $\dot\rho$ as $(ax)_{,t}\dot t+(ax)_{,x}\dot x$, use the 
relation
\eqn{crutch}{1-\frac{x^2}{R^2}=\frac{(ax)_{,x}^2\dot x^2}{\dot t^2-1} 
}
once, assemble a perfect square from the pieces, and then use 
(\ref{crutch}) again to end up with the first piece in (\ref{mixed}). 
Equation (\ref{crutch}) here is just a rewriting of (\ref{peeb-match1}). 
An 
analogous transformation can be done for the open-FRW part of the 
equation.

The advantage of the form (\ref{standard-junc}) is that now we can use the 
LTB
equation of motion (resp. the open-FRW equation of motion) to simplify the
two terms further to write
\begin{eqnarray}
&&\Delta_{out}=-1+\Big(\frac{A}{a^3}+\frac{\Lambda_{out}}{3}\Big)\rho^2,
\label{delta+}\\
&&\Delta_{in}=-1+\frac{\Lambda_{in}}{3}\rho^2\label{delta-}.
\end{eqnarray}

Now, we take up the task of writing these equations purely in the LTB
coordinates, for otherwise, despite being correct, they will be of no
practical value in computations because of the mixing of coordinates.
Using (\ref{delta+}) and (\ref{delta-}), we can
write (\ref{standard-junc}) as
\eqn{rhodot}{{\dot \rho}^2=\rho^2 B^2-1,}where
\eqn{B}{B^2=\frac{\Lambda_{in}}{3}+\Big(\frac{\sigma}{4}+\frac{1}{\sigma}
\Big( \frac{\Lambda_{out}-\Lambda_{in}}{3}+\frac{A}{^3}\Big)\Big)^2.
}
This form is useful because we just have to focus on
${\dot \rho}$ and
$\rho$ because everything else is already manifestly in
LTB coordinates.
The idea is that instead of choosing
$\rho(\tau)$ as the curve for the
bubble-evolution,
we want to
parameterize it as $x(t)$. Since $x$ and $t$ are dependent
variables {\em{on}} the shell, this is legitimate. So we write
$\rho=ax$, and rewrite ${\dot \rho}$ as
\eqn{rhodot-xt}{{\dot \rho}=\frac{\Big(x\partial_t
a+(ax),_{x}\frac{dx}{dt}\Big)}{\sqrt{1-\frac{{(ax),_x}^2}{1-x^2/R^2}
\Big(\frac{dx}{dt}\Big)^2}}.}
Using this in (\ref{rhodot}), and using the LTB equation of motion for
$\partial_t a$, we end up with,
\eqn{eomforshell1}{\frac{dx}{dt}=\frac{\left\{\begin{array}{cc}\hspace{-2.25in}-x(1-\frac{x^2}{R^2})\Big(\frac{A}{a}+\frac{\Lambda
a^2}{3}-
\frac{1}{R^2}\Big)^{1/2} \pm\\
\hspace{1in}{\pm}x\Big\{(1-\frac{x^2}{R^2})(a^2x^2B^2-1)\Big(a^2B^2-\frac{A}{a}-\frac{\Lambda
a^2}{3}\Big)\Big\}^{1/2} \end{array}\right\}}
{(ax),_{x}(a^2x^2B^2-x^2/R^2)}}
This is the first of the two shell evolution equations (the other being
(\ref{Ton})) in a form that is
directly applicable
for numerical simulations. The  sign on the square root comes from
the square root of (\ref{rhodot}) and has to be chosen according to
whether
the shell is expanding or contracting.

The second of the coupled shell-equations is easily translated into the
LTB coordinates as well. On the LTB side, matter takes the form
dust + cosmological constant (\ref{EOM}). So, using (\ref{peeble-normal}),
\eqn{Ttn+}{T^n_{\tau (out)}\equiv e^t_\tau T^t_t n_t+e^x_\tau T^x_x n_x
=-\frac{\gamma_{out}(ax)_{,x}\frac{dx}{dt}d(x,t)\sqrt{1-x^2/R^2}}
{1-x^2/R^2-(ax)_{,x}^2(dx/dt)^2}.}
On the inside, the only matter is the vacuum energy, which gives rise to
$T_{\tau (in)}^n=0$. These along with (\ref{peeb-match1}) can be used to
write (\ref{Ton}) in the form
\eqn{eomforshell2}{\frac{d\sigma}{dt}+\frac{\Big(x\partial_t
a+(ax),_{x}\frac{dx}{dt}\Big)}{ax}(\sigma+P)-\frac{\gamma_{out}(ax)_{,x}\frac{dx}{dt}d(x,t)}
{\sqrt{1-\frac{x^2}{R^2}-(ax)_{,x}^2(dx/dt)^2}}=0.}
Together with an equation of state relating $\sigma$ to $P$, and the
LTB
equation of motion (\ref{EOM}), 
(\ref{eomforshell1}, \ref{eomforshell2}) comprise the starting point
for the
various special cases we study in the main text of this paper.

When solving for shell dynamics, we should demand that the
positive energy condition be satisfied, that is
\eqn{pec}{ \sigma > 0. }
Using the formulas derived here (in particular
(\ref{standard-junc})), we can write :
\eqn{deltas} {\Delta_{out} - \Delta_{in} = \frac{\rho^2 \sigma^2}{4} +
\gamma_{out} \rho \sigma ( \dot\rho^2 - \Delta_{out} )^{1/2}, }
which together with (\ref{pec}) implies:
$$ \Delta_{out} - \Delta_{in}  >  \frac{\rho^2 \sigma^2}{4} \hspace{4ex}  
\mbox{if}  \hspace{4ex} \gamma_{out} = +1 $$
$$ \Delta_{out} - \Delta_{in}  <  \frac{\rho^2 \sigma^2}{4} \hspace{4ex}  
\mbox{if}  \hspace{4ex} \gamma_{out} = -1 $$
Furthermore, defining:

\eqn{def_xi} {\xi \equiv \frac {4 (\Delta_{out} - \Delta_{in} )} {\rho^2 
\sigma^2} }
and using (\ref{standard-junc}), the following relationship holds:

\eqn{xi_constraint} { \gamma_{in} \mid \xi + 1 \mid  - \gamma_{out} \mid 
\xi - 1 \mid = 2 }

Depending on the outer and inner geometry, and given the energy densities
both inside and outside the bubble, there will only be a certain range of 
possible
values for the energy density $\sigma$ on the surface of the bubble 
consistent with (\ref{xi_constraint}). We will use this to 
constrain the value of $\sigma$ in the various cases addressed in this 
paper.

%


\end{document}